\documentclass[prd,reprint,nofootinbib,showpacs,aps]{revtex4-1}
\usepackage{amsmath, amssymb,bm,xcolor,graphicx,mathtools,enumitem}
\usepackage[colorlinks=true,citecolor=blue]{hyperref}
\usepackage[T1]{fontenc}
\usepackage{soul}
\setstcolor{red}

\newcommand{\ret}{\mathrm{ret}}
\newcommand{\adv}{\mathrm{adv}}

\newcommand{\subE}{ \textrm{\fontsize{6}{8}\selectfont{E}} }

\newcommand{\subB}{ \textrm{\fontsize{6}{8}\selectfont{B}} }
\newcommand{\subG}{ \textrm{\fontsize{6}{8}\selectfont{G}} }

\newcommand{\subS}{\textrm{\fontsize{6}{8}\selectfont{S}}}
\newcommand{\subN}{\textrm{\fontsize{6}{8}\selectfont{N}}}
\newcommand{\subEM}{\textrm{\fontsize{6}{8}\selectfont{EM}}}
\newcommand{\subDW}{\textrm{\fontsize{6}{8}\selectfont{DW}}}

\newcommand\redsout{\bgroup\markoverwith{\textcolor{red}{\rule[0.5ex]{2pt}{0.4pt}}}\ULon}

\def\be{\begin{equation}}
\def\ee{\end{equation}}

\begin{document}
\title{Foundations of the self-force problem in arbitrary dimensions}

\author{Abraham I. Harte$^1$, Peter Taylor$^{1}$ and \'Eanna \'E. Flanagan$^{2,3}$}
\affiliation{$^1$Centre for Astrophysics and Relativity, School of Mathematical Sciences\\ Dublin City University,
Glasnevin, Dublin 9, Ireland}
\affiliation{$^2$Cornell Center for Astrophysics and Planetary Science, Cornell
University, Ithaca, NY 14853}
\affiliation{$^3$Department of Physics, Cornell University, Ithaca, NY 14853}

% ----------------------------------------------------------------------
%
% TIME OF DAY
%
\newcount\hh
\newcount\mm
\mm=\time
\hh=\time
\divide\hh by 60
\divide\mm by 60
\multiply\mm by 60
\mm=-\mm
\advance\mm by \time
\def\hhmm{\number\hh:\ifnum\mm<10{}0\fi\number\mm}
% ----------------------------------------------------------------------
%\date{ {\color{red} Draft version 42 of 20 February, 2016; printed \today{} at \hhmm}}

\begin{abstract}
The self-force problem---which asks how self-interaction affects a body's motion---has been poorly studied for spacetime dimensions $d \neq 4$. We remedy this for all $d \geq 3$ by nonperturbatively constructing momenta such that forces and torques acting on extended, self-interacting electromagnetic charges have the same functional forms as their test body counterparts. The electromagnetic field which appears in the resulting laws of motion is not however the physical one, but a certain effective surrogate which we derive. For even $d\ge 4$, explicit momenta are identified such that this surrogate field satisfies the source-free Maxwell equations; laws of motion in these cases can be obtained similarly to those in the well-known four-dimensional Detweiler-Whiting prescription. For odd $d$, no analog of the Detweiler-Whiting prescription exists. Nevertheless, we derive its replacement. These general results are used to obtain explicit point-particle self-forces and self-torques in Minkowski spacetimes with various dimensions. Among various characteristics of the resulting equations, perhaps the most arresting is that an initially-stationary charge which is briefly kicked in $2+1$ dimensions asymptotically returns to rest.
\end{abstract}
\vskip 1pc

\maketitle
\section{Introduction}

The motion of small bodies is central to some of the most enduring problems in physics. If such a body is coupled to an electromagnetic, gravitational, or other long-range field, it may be subject to net forces exerted by its own contributions to that field. This ``self-force'' strongly influences, for example, charged particles circulating in particle accelerators and the shrinking orbits of black hole binaries due to the emission of gravitational radiation. However, the apparent simplicity of the statement of the self-force problem belies a number of physical and mathematical subtleties. This has led to more than a century of literature on the subject; see \cite{Thomson, Abraham, Lorentz, Dirac, DeWittBrehme, Nodvik, Spohn, Yaghjian, GrallaHarteWald, HarteEM} for some electromagnetic examples.

While motivations for working on the self-force problem have varied considerably over the years, the past two decades have seen a concerted effort---motivated largely by gravitational wave astronomy---to understand the gravitational self-force problem in general relativity. This has led to a number of theoretical and computational advances which considerably improve our understanding of classical self-interaction, in both the gravitational and electromagnetic contexts \cite{PoissonLR, HarteReview, PoundReview, WardellReview}. Separately, new aspects of the electromagnetic self-force are beginning to be accessible to investigation via high-power laser experiments \cite{laserRev, Laser2}.

In this paper, we move beyond the existing literature, which almost exclusively focuses on four spacetime dimensions, to rigorously study self-interaction in all dimensions $d \geq 3$. There are several reasons for this: First, considerations in different numbers of dimensions refine our understanding of precisely what is important and what is not; lessons learned in this way may significantly inform future considerations even in four dimensions, particularly in more complicated theories which have not yet been understood. Second, considerations of theories in non-physical numbers of dimensions can, via holographic dualities, be related to ordinary four-dimensional systems; for example, the five-dimensional self-force might be used to understand jet quenching in four-dimensional quark-gluon plasmas \cite{Shuryak2}. Third, self-forces in odd numbers of spacetime dimensions are qualitatively very different from those in even dimensions. In particular, one crucial ingredient of the self-force framework is the derivation of an appropriate map from the physical field to an effective surrogate in which, e.g., the Lorentz force law is preserved despite the possible presence of radiation reaction and similar effects; this map had not been previously understood in odd dimensions and differs considerably from its even-dimensional counterpart. Despite these differences, a significant portion of this paper is devoted to developing a unified formalism which applies for any parity of dimension. Our results subsume the Detweiler-Whiting scheme which was originally given in four dimensions \cite{DetweilerWhiting2003, PoissonLR, HarteReview}.

Our final reason for considering different numbers of dimensions is that the self-force in three spacetime dimensions may be proportionally stronger than in four dimensions, both in terms of instantaneous magnitude \cite{HFTPaper1} and---as argued below---in the particularly slow decay of fields in this context. The latter property implies that self-interaction encodes a strong ``memory'' of a system's past. Moreover, systems in which this is relevant may be accessible to experiment. For example, ``pilot wave hydrodynamics'' involves a number of striking phenomena observed to be associated with oil droplets bouncing on a vibrating bath \cite{PilotWave}. Each bounce generates surface waves on the bath, but these waves also affect the horizontal motion of the droplet. This type of feedback with a long-range field (the surface waves) is reminiscent of a self-force problem in two spatial dimensions. There are also a variety of condensed matter systems which act as though they are confined to one or two spatial dimensions \cite{Superfluid, TopIns}, and some of our considerations may be relevant there as well.

We do not consider any particular fluid or condensed matter system in this paper, but instead explore a standard electromagnetic self-force problem in different numbers of dimensions. To the best of our knowledge, the literature does not contain any rigorous derivations of the self-force other than in four dimensions, except for recent work restricted to static bodies \cite{HFTPaper1} (see however work on the radiation reaction component of the self-force in effective field theory \cite{Ofek1, Ofek2}). The fully-dynamical self-force considered here is considerably different and more rich than its static counterpart. We choose to focus on the standard electromagnetic self-force problem purely for clarity of exposition; our results generalize to extended bodies coupled to Klein-Gordon fields and to linearized gravitational fields satisfying the Einstein equations. There are also no conceptual obstacles to considering non-trivial boundary conditions such as those appropriate for describing analogs of the pilot-wave hydrodynamics experiments mentioned above.

A layout of the paper is as follows: In Section \ref{Sect:NonPert}, we
briefly review, and then apply and extend, a non-perturbative
formalism \cite{HarteScalar, HarteEM, HarteGrav, HarteReview} which
provides a general framework for the problem of motion of strongly
self-interacting extended bodies. Not all of the intricate details of
this formalism are required to absorb the essential points of this
paper, although some of the most relevant aspects are recounted here. We apply
them to derive equations of motion for an extended body coupled to an
electromagnetic field in arbitrary dimensions, including all
self-interaction effects. The resultant equations are structurally
identical to extended test-body equations, except that the physical field
in the test-body equations is mapped to an effective field which encapsulates all self-force and self-torque effects. The cost of this map is
that the stress-energy tensor of the body is renormalized as it
appears in the laws of motion. This renormalization is
well-controlled, however, being both finite and quasi-local (in a
sense we make precise later). Moreover, our laws of motion admit
well-defined point-particle limits. In Section \ref{Sect:Flat}, we
discuss these point-particle limits with retarded boundary conditions in Minkowski spacetimes with various dimensions, obtaining explicit
point-particle self-forces and self-torques in these cases. In
Section \ref{Sect:pheno}, we discuss some interesting phenomenology
that occurs in odd numbers of spacetime dimensions. Some of the issues
that arise here remain open problems which we hope will inspire
further interest. We particularly focus on self-force phenomenology in
$d=3$, where self-interaction effects can be relatively large and
where there is a very strong history dependence. One particularly
striking example of the latter shows up in the case of a charge which
is initially stationary and is then given a kick by an
externally-imposed force. Our analysis shows that the
slowly-decaying fields in this case cause such a particle to return to
rest at late times, a phenomenon reminiscent of Aristotelian
physics.

Throughout this paper, units are chosen in which $G=c=1$, the metric signature is positive, abstract indices are denoted by $a$, $b$, $\dots$, spacetime coordinate indices by $\mu$, $\nu$, $\dots$, and spatial coordinate indices by $i$, $j$, $\dots$.

\section{Extended bodies and non-perturbative laws of motion}
\label{Sect:NonPert}

Our strategy is not to obtain a ``point particle self-force'' as any kind of fundamental concept, but instead to derive laws of motion first for extended charge distributions and then to evaluate point-particle limits of those laws. Although we focus for concreteness on the electromagnetic self-force problem, analogous results are easily obtained for the scalar and (at least the first order) gravitational self-force. The approach adopted here is based on a nonperturbative formalism developed by one of us \cite{HarteScalar, HarteEM, HarteGrav, HarteReview}, which provides a rigorous framework with which to analyze problems of motion in a wide variety of contexts. Crucially, most of this framework is agnostic to the number of spacetime dimensions.

\subsection{Preliminaries}

In the electromagnetic context, a finite extended body in a $d$-dimensional spacetime $(M,g_{ab})$ is associated with a nonsingular conserved current density $J_a$ and a nonsingular stress-energy tensor $T^{ab}_\subB = T^{(ab)}_\subB$. The support of $T^{ab}_\subB$ may be identified with the body's worldtube $\mathfrak{W} \subset M$, and that of $J_a$ is assumed not to extend beyond $\mathfrak{W}$. Furthermore, we suppose that the body's worldtube is spatially compact and that the electromagnetic field $F_{ab}$ satisfies  Maxwell's equations
\begin{equation}
        \nabla_{[a} F_{bc]} = 0, \qquad \nabla^b F_{ab} = \omega_{d-1} J_a,
        \label{Maxwell}
\end{equation}
in a neighborhood of $\mathfrak{W}$, where
\begin{equation}
        \omega_{d-1} \equiv \frac{2 \pi^{\frac{1}{2}(d-1)}}{\Gamma(\frac{1}{2}(d-1))}
\end{equation}
is equal to the area of a unit sphere in $\mathbb{R}^{d-1}$. The dynamical evolution of such an extended body may be understood, at least in part, via energy and momentum exchanges between that body and the electromagnetic field. Such exchanges are more precisely described by the conservation of the system's total stress-energy tensor $T^{ab} = T^{(ab)}$, in the sense that
\begin{equation}
        \nabla_b T^{ab} = 0.
        \label{TCons}
\end{equation}
Although various arguments can be made for how to ``most naturally'' split $T^{ab}$ into electromagnetic and material components inside of a body, particularly if that body possesses nontrivial dielectric or related properties \cite{AbrahamMinkowski}, we pragmatically extend the vacuum expression
\begin{equation}
        T^{ab}_\subEM \equiv \frac{1}{\omega_{d-1}} \left( F^{a}{}_{c} F^{bc} - \frac{1}{4} g^{ab}  F^{cd} F_{cd} \right)
        \label{Tem}
\end{equation}
for the electromagnetic stress-energy tensor into the interior of $\mathfrak{W}$: $T^{ab} = T^{ab}_\mathrm{B} + T^{ab}_\subEM$ everywhere of interest. This sum may in fact be viewed as a definition for $T^{ab}_\subB$ in terms of $T^{ab}$ and $F_{ab}$. Adopting it, \eqref{Maxwell} and \eqref{TCons} imply that
\begin{equation}
        \nabla_b T^{ab}_\subB = F^{a}{}_{b} J^b.
        \label{TConsBody}
\end{equation}
Every portion of an extended charge is thus acted upon by the Lorentz force density $F_{ab} J^b$.

The question we now ask is how this force density ``integrates up'' to affect a body's overall motion. One difficulty is that Maxwell's equations imply that $F_{ab} J^b$ depends nonlinearly and nonlocally on $J_a$, and can be almost arbitrarily complicated. Despite this, experience suggests that there are physically-interesting regimes in which the (appropriately-defined) \textit{net} force is not complicated at all: Laws of motion arise in which net forces involve a body's internal structure only via its first nonvanishing multipole moments. That these laws do not depend on finer details of $J_a$ lends them a certain degree of ``universality.'' Deriving this universality and making it precise is the essence of the self-force problem.

We emphasize that it is only certain ``bulk'' features of an extended body which can be described as universal. Individual objects may develop internal oscillations, turbulent flows, or other fine details which strongly depend on material composition, thermodynamic effects, and other characteristics which are difficult both to specify and to model. As in Newtonian celestial mechanics, our goal here is to ignore as many of these details as possible, and to instead identify certain features of $T^{ab}_\subB$ and $J_a$ which i) describe a body's behavior ``as a whole,'' and ii) whose evolution is only weakly coupled to a body's internal details. In Newtonian celestial mechanics, these two properties are well-known to hold for the linear and angular momenta associated with each celestial body. We now generalize the concept of momentum to describe relativistic extended bodies coupled to electromagnetic fields in $d$-dimensional spacetimes.

\subsection{Momentum}

Although there is a strong physical expectation that some momentum-like quantity can be defined for extended, relativistic charge distributions, non-pathological definitions are not so easily written down. The problem is particularly acute for bodies with significant self-fields, essentially because, i) those self-fields carry energy and momentum, ii) they rearrange themselves to adjust to any motions associated with $\mathfrak{W}$, and iii) self-fields may extend far outside a body's material boundaries. The first and second of these points suggest that a body's self-field contributes, e.g., to its inertia---a fact already recognized by the end of the 19th century \cite{Thomson}. Indeed, four-dimensional electromagnetic self-fields are now known to contribute to (or ``renormalize'') not only a body's apparent mass, but all of its stress-energy tensor \cite{HarteTrenorm, HarteEM, HarteReview}.

While there must be some sense in which these effects generalize to any number of dimensions, they are nontrivial to compute even when $d=4$. Fundamentally, this is because it is difficult to guess a suitable definition for the momentum which treats a body's material and self-field components as one. Consider, for example, a ``renormalized body momentum'' defined by appropriately integrating $T^{ab}_\subB + T^{ab}_\mathrm{self}$, where $T^{ab}_\mathrm{self}$ is an electromagnetic stress-energy tensor which is quadratic in a suitably-defined self-field. Any ``mass'' associated with such a definition would clearly depend on properties of the self-field at arbitrarily-large distances, and hence on the state of system over arbitarily-long times. It would not describe a body's instantaneous resistance to applied forces, and would thus be a poor definition of mass. Indeed, the degree of nonlocality associated with naive momentum definitions such as this renders them physically unacceptable, at least in a non-perturbative context. While the problem is considerably alleviated at low orders in the perturbative expansions commonly associated with point-particle limits \cite{GrallaHarteWald, MoxonFlanagan}, it is difficult to extend the methods employed in those contexts to higher orders in perturbation theory, or even to apply them at low orders when Huygens' principle is violated.

The nonperturbative formalism \cite{HarteReview} employed here affords a different approach, allowing us to derive---rather than postulate---physically-acceptable definitions for the linear and angular momenta of a charged extended body. Our results do not depend on any limiting process, and are nonlocal only in the sense that they depend on the state of the system over spatial and temporal scales comparable to a suitably-defined diameter for $\mathfrak{W}$. No smaller degree of nonlocality could reasonably be expected, even for a body with no self-field whatsoever. Schematically, we obtain this definition by first proposing a ``bare momentum,'' essentially a guess which need not take into account any self-field effects, and then deriving a correction to that definition which maintains locality while also decoupling forces and torques from a body's internal details.

Our momenta depend on a choice of origin, which we take to be a timelike worldline parametrized by $\gamma(\tau)$. Such origins are required even in Newtonian mechanics at least to define the angular momentum, and it is only the global parallelism of Euclidean space which prevents them from also being necessary to define the Newtonian linear momentum. Global parallelism cannot be assumed here, so an origin is needed for both our linear and angular momenta. We additionally need to specify a family of spacelike hypersurfaces $\mathfrak{B}_\tau \ni \gamma(\tau)$ which foliate $\mathfrak{W}$, thus fixing a notion of time inside a body's worldtube. Specific worldlines and specific hypersurfaces may be fixed using spin supplementary and similar conditions; see Section \ref{Sect:LawsofMotion}. At this stage, however, we assume only that some prescription has been given. Its details do not matter. Next, we define the bare ``generalized momentum'' $P_\tau [ \cdot ]$ at time $\tau$ via
\begin{align}
        P_\tau [\xi_a] \equiv \int_{\mathfrak{B}_\tau} \! dS_a \Bigg[ T_\subB^{ab}(x) \xi_b(x) + J^a(x) \int_0^1 du u^{-1}
        \nonumber
        \\
        ~  \times \nabla_{b'} \sigma (y'(u), \gamma(\tau) ) F^{b'c'} ( y'(u) ) \xi_{c'} ( y'(u)) \Bigg],
        \label{PDef}
\end{align}
where $\sigma(x,x')$ denotes Synge's world function, defined to equal one half of the squared geodesic distance between its arguments, $y'(u)$ is an affinely-parametrized geodesic for which $y'(0) = \gamma(\tau)$ and $y'(1) = x$, and $\xi^a(x)$ is any vector field drawn from a certain finite-dimensional vector space referred to as the space of generalized Killing fields (GKFs). The GKFs are defined more precisely in \cite{HarteSyms, HarteReview}, and coincide with the space of all ordinary Killing vector fields in maximally-symmetric spacetimes. More generally, the GKFs always form a vector space with dimension $\frac{1}{2} d(d+1)$. The generalized momentum at fixed time is a linear operator on this space, and may therefore be viewed as a $\frac{1}{2} d(d+1)$-dimensional vector in the vector space dual to the space of GKFs. It simultaneously encodes both the linear and angular momentum of an extended body. Just as electric and magnetic fields are best thought of as two aspects of a single electromagnetic field, a body's linear and angular momenta are two aspects of a single more-fundamental structure: the generalized momentum.

More precisely, a bare linear momentum $p_a(\tau)$ and a bare angular momentum $S_{ab} = S_{[ab]} (\tau)$ may be implicitly defined by combining \eqref{PDef} with
\begin{equation}
        P_\tau[\xi_a] \equiv p_a(\tau) \xi^a ( \gamma(\tau) ) + \frac{1}{2} S^{ab} (\tau) \nabla_a \xi_b ( \gamma(\tau) ).
        \label{pSDef}
\end{equation}
$P_\tau[\xi_a]$ thus returns a linear combination of the linear and angular momenta. The particular choice of $\xi^a$ controls which linear combination is obtained. If $\psi^a$ is, e.g., a translational Killing field in flat spacetime, $P_\tau[\psi_a]$ returns the component of linear momentum associated with that translation. Regardless, varying over all possible GKFs results in integral formulae for $p_a$ and $S_{ab}$ which involve $T^{ab}_\subB$, $J_a$, and $F_{ab}$. These formulae coincide with definitions originally proposed by Dixon \cite{Dix67, Dix70a, Dix74}, who sought multipole moments for $T^{ab}_\subB$ in which stress-energy conservation implies differential constraints \textit{only} on the monopole and dipole moments. That goal was achieved, meaning that there is a sense in which ordinary differential equations with the form $\dot{p}_a = (\ldots)$ and $\dot{S}_{ab} = (\ldots)$ are fully equivalent to the partial differential equation \eqref{TConsBody}. Regardless, these definitions for the momenta reduce to textbook ones \cite{MTW} in flat spacetime and with vanishing electromagnetic fields. They also give rise, more generally, to simple conservation laws whenever there exists a Killing vector field which is also a symmetry of $F_{ab}$ \cite{Dix70a, HarteReview}. We note that this last property would fail to hold if the (less familiar) electromagnetic terms in \eqref{PDef} were omitted.

Now that a bare momentum has been proposed, its evolution may be understood by differentiating \eqref{PDef} with respect to $\tau$ while applying \eqref{TConsBody}. The resulting rate of change may be interpreted as encoding bare forces and bare torques, which can again be written as integrals over a body's interior:
\begin{equation}
        \frac{d}{d\tau} P_\tau[\xi_a] = \mathcal{F}^\subG_\tau[ T^{ab}_\subB; \xi_c] + \mathcal{F}^\subEM_\tau[ F_{ab}, J_c ; \xi_d ],
        \label{PDotGen}
\end{equation}
where the gravitational generalized force is given by the bilinear functional
\begin{align}
        \mathcal{F}^\subG_\tau[ T^{ab}_\subB; \xi_c] = \frac{1}{2} \int_{\mathfrak{B}_\tau} \! \! dS T^{ab}_\subB \mathcal{L}_\xi g_{ab} ,
\end{align}
and the electromagnetic generalized force by the trilinear functional
\begin{align}
        \mathcal{F}^\subEM_\tau[ F_{ab}, J_c ; \xi_d ]= \int_{\mathfrak{B}_\tau} \!\! dS J^b \bigg[  \xi^a F_{ab} + \nabla_b \int_0^1 du u^{-1}
        \nonumber
        \\
        ~ \times \nabla_{b'} \sigma(y'(u), \gamma(\tau)) F^{b'c'}(y'(u)) \xi_{c'}(y'(u)) \bigg].
\end{align}
The $\xi^a$ here again denote GKFs and the $\mathcal{L}_\xi$ are Lie derivatives. Any particular GKF which is substituted into these equations may be viewed as selecting the corresponding component of the generalized gravitational and electromagnetic force vectors.

If one has full knowledge of $T^{ab}_\subB$ and $F_{ab}$, and hence $J_a = \omega^{-1}_{d-1} \nabla^b F_{ab}$, electromagnetic forces and torques can be computed by directly evaluating the integral $\mathcal{F}^\subEM_\tau[ F_{ab}, J_c ; \xi_d ]$ for all possible GKFs. However, there is little point to such descriptions if the system is already understood in full detail. Momenta and related concepts are most valuable precisely when a system's details are only partially specified, or alternatively, if one would like to understand a class of distinct systems which nevertheless share some bulk features.

With this context in mind, forces and torques associated with the definition \eqref{PDef} may be shown to have certain undesirable characteristics. First, unless self-fields are negligible, the functionals $\mathcal{F}^{\subEM}_\tau[ F_{ab}, J_c ; \xi_d ]$ are difficult to approximate as-is. In particular, it is not clear how to evaluate them if, e.g., only the first few multipole moments of $J_a$ are known. More precisely, it is the actual self-field which must be negligible in order to make reasonable approximations, and not only some suitable integral of that self-field. There are very few realistic settings in which such an approximation can be justified, particularly if smallness is also demanded for the derivatives of a body's self-field. In some cases, this difficulty is merely a question of calculational practicality. In others, it is essential: There are important examples in which bare forces and torques may be directly shown to depend on the detailed nature of $J_a$. When this occurs, the laws of motion associated with $p_a$ and $S_{ab}$ cannot be deemed universal. Nevertheless, we find closely-related momenta which do obey universal laws of motion.

The formalism we employ accomplishes this by providing a set of tools which allow one to easily establish identities with the form
\begin{align}
        \mathcal{F}^\subEM_\tau[ F_{ab}, J_c ; \xi_d ] = \mathcal{F}^\subEM_\tau[ \hat{F}_{ab}, J_c ; \xi_d ]- \frac{d}{d\tau} \delta P_\tau[J_a;\xi_b]
        \nonumber
        \\
        ~  + \mathcal{F}^\subG_\tau[ \delta T^{ab}_\subB [J_c] ; \xi_d] ,
        \label{fVary}
\end{align}
for a wide variety of nonlocal field transformations $F_{ab} \mapsto \hat{F}_{ab}$. The nature of this field transformation determines specific forms for the functionals $\delta T^{ab}_{\subB} [ J_a ]$ and $\delta P_\tau [J_a;\xi_b]$, both of which are nonlinear in $J_a$. The point of \eqref{fVary} is to describe how electromagnetic forces and torques exerted on a charge distribution $J_a$ by a field $F_{ab}$ can be computed in terms of forces and torques exerted by a potentially-simpler ``effective field'' $\hat{F}_{ab}[F_{cd}]$. Substituting \eqref{fVary} into \eqref{PDotGen} suggests the definition
\begin{equation}
        \hat{P}_\tau [ \xi_a ] \equiv P_\tau [ \xi_a ] + \delta P_\tau [J_a; \xi_b]
\end{equation}
for the renormalized momentum, which is seen to satisfy
\begin{equation}
        \frac{d}{d\tau} \hat{P}_\tau [ \xi_a ] = \mathcal{F}^\subG_\tau[ \hat{T}^{ab}_\subB ; \xi_c] +  \mathcal{F}_\tau^\subEM [ \hat{F}_{ab}, J_c;\xi_d ].
        \label{PDotGenHat}
\end{equation}
Here, $\hat{T}^{ab}_\subB \equiv T^{ab}_\subB + \delta T^{ab}_\subB$ and the dependence of $\hat{P}_\tau$ on $T^{ab}_\subB$ and $J_a$ has been suppressed. It follows that if the map $F_{ab} \mapsto \hat{F}_{ab}$ satisfies appropriate conditions---to be discussed in more detail below---electromagnetic forces and torques appear to be exerted not by the physical field $F_{ab}$, but by a particular surrogate $\hat{F}_{ab}$. The cost for this replacement is effectively a renormalization of the body's apparent stress-energy tensor, which affects its apparent momenta and the couplings which appear in gravitational forces and torques. Both of these effects can be physically interpreted as contributions due to the body's self-field. Note that although $T^{ab}_\subB$ is renormalized here, $J_a$ is not, essentially because Maxwell's equations are linear.

Eq. \eqref{PDotGenHat} has an advantage over \eqref{PDotGen} when $\hat{F}_{ab}$ behaves more simply inside a body than $F_{ab}$. More specifically, there should be a wider variety of circumstances in which a well-chosen $\hat{F}_{ab}$ varies slowly throughout each $\mathfrak{B}_\tau$. Whenever this occurs, useful multipole expansions may be found for $\mathcal{F}_\tau^\subEM [ \hat{F}_{ab}, J_c;\xi_d ]$, resulting in electromagnetic forces and torques which are identical in form to standard test body expressions, but with all fields in those expressions equal to $\hat{F}_{ab}$. Similarly, a multipole expansion for $\mathcal{F}^\subG_\tau[ \hat{T}^{ab}_\subB ; \xi_d]$ results in standard expressions for the gravitational forces and torques acting on an extended test body, but with stress-energy multipole moments which are somewhat different from those which might have been computed using $T^{ab}_\subB$ alone. These gravitational effects involve only quadrupole and higher order moments, and vanish entirely in maximally-symmetric spacetimes.

\subsection{Effective test bodies and effective fields}

Laws of motion which are structurally identical to those satisfied by test bodies moving in an effective fictitious field are used today to organize essentially all known $d=4$ self-force results---whether in electromagnetism, scalar field theory, or general relativity \cite{DetweilerWhiting2003, PoissonLR, HarteReview}. Such principles have also been employed to understand \textit{static} scalar and electromagnetic self-interaction for more general values of $d$ \cite{HFTPaper1}. Indeed, we may view the motion of a self-interacting body moving in a certain physical field as ``equivalent'' to the motion of an ``effective test body'' in an appropriate effective field.

It is instructive to note that a particularly simple principle of this sort holds even in Newtonian gravity \cite{HarteScalar, HarteReview}, where it provides the foundation for celestial mechanics. Using standard definitions for the linear and angular momenta of a Newtonian extended body with mass density $\rho$ coupled to a gravitational potential $\phi$, the Newtonian generalized force at time $\tau$ may be shown to be
\begin{align}
        \frac{d}{d\tau} P^\subN_\tau[\xi_a] &= \mathcal{F}^{\subN}_\tau[\phi,\rho; \xi_a]
        \nonumber
        \\
        & = - \int_{\mathfrak{B}_\tau} \! \! \rho(\bm{x},\tau) \mathcal{L}_\xi \phi(\bm{x},\tau) d^3 \bm{x},
        \label{FNewt}
\end{align}
where $\mathfrak{B}_\tau \subset \mathbb{R}^3$ denotes the body's location at time $\tau$, $\xi^a$ is any Euclidean Killing vector field, and $\mathcal{L}_\xi$ again denotes the Lie derivative with respect to $\xi^a$. As in electromagnetism, this generalized force is a trilinear functional of the potential, its source, and a vector field. Now, it is straightforward to show that if $G(\bm{x},\bm{x}') = G(\bm{x}',\bm{x})$ and $\mathcal{L}_\xi G(\bm{x},\bm{x}') = 0$ for all Killing vector fields $\xi^a$, and if
\begin{equation}
        \hat{\phi}(\bm{x} ,\tau ) \equiv \phi(\bm{x},\tau) - \int_{\mathfrak{B}_\tau} \!\! \rho(\bm{x}',\tau) G(\bm{x},\bm{x}') d^3\bm{x}',
        \label{phiHat}
\end{equation}
forces and torques computed using $\phi$ must be identical to those computed using $\hat{\phi}$:
\begin{equation}
        \mathcal{F}^{\subN}_\tau[\phi,\rho; \xi_a] = \mathcal{F}^{\subN}_\tau[\hat{\phi},\rho; \xi_a].
        \label{fVaryNewt}
\end{equation}
This result is directly analogous to \eqref{fVary}. Although the language used here is unconventional, the conclusion is not; it is standard to apply a result equivalent to \eqref{fVaryNewt} specialized so that $G(\bm{x},\bm{x}') = - |\bm{x} - \bm{x}'|^{-1}$. In that context, $G(\bm{x}, \bm{x}')$ is a Green function for the Laplace equation and $\hat{\phi}$ satisfies the source-free field equation $\nabla^2 \hat{\phi} = 0$. Indeed, the effective field here is just the external potential. Given \eqref{phiHat} and the Newtonian field equation $\rho = \nabla^2 \phi/4\pi$, the external potential may be viewed as a nonlocal linear functional of the physical potential $\phi$.

The freedom to compute forces using $\hat{\phi}$ instead of $\phi$ is essential for the development of useful approximations. For example, if $\mathcal{F}^\subN_\tau[\hat{\phi},\rho;\xi_a]$ is evaluated for a body in which $\hat{\phi}$ does not vary too much throughout $\mathfrak{B}_\tau$, it is straightforward to recover the usual gravitational force $- m \nabla_a \hat{\phi}$ known to act on a massive body in Newtonian mechanics. The superficially-similar expression $- m \nabla_a \phi$ would arise if $\phi$ varied slowly as well, although this is rarely the case. Indeed, the former approximation for the force is valid much more generally than the latter. This difference is most striking in a point-particle limit, wherein $-m\nabla_a \hat{\phi}$ remains a valid approximation while $-m\nabla_a \phi$ is not even computable. This type of improvement persists also at higher multipole orders, and the usual laws of motion in Newtonian celestial mechanics are written as laws of motion associated with (possibly extended) test bodies moving in the external potential, not the physical one; even in Newtonian gravity, the effects of self-fields must be properly understood before obtaining useful laws of motion.

Our goal here is to extend these ideas to fully-dynamical electromagnetically-interacting systems in all dimensions $d \geq 3$. Much of the content of the general principle that self-interacting bodies move like effective test bodies is embedded in the precise specifications for the renormalized momenta and the effective fields for which the statement is true, so this perspective suggests that the problem of motion can be solved by finding ``appropriate'' momenta and effective fields. Doing so turns out to be possible much more generally and simply than the older, more-explicit approach to the self-force problem, where forces were directly computed using some specific definition for the momentum, some specific approximation scheme, specific boundary conditions, and so on.

We now search for a nonlocal transformation $F_{ab} \mapsto \hat{F}_{ab}[F_{cd}]$ such that i) Eq. \eqref{fVary} holds, ii) all renormalizations implicit in that equation are physically acceptable, and iii) the transformed field has an ``external character'' similar to that of the Newtonian external field. We do so by first using the Newtonian field transformation \eqref{phiHat} as a model and defining the effective electromagnetic field via
\begin{equation}
        \hat{F}_{ab}(x) \equiv F_{ab} (x) - 2 \int  \nabla_{[a} G_{a]a'}(x,x') J^{a'}(x') dV',
        \label{Fhat}
\end{equation}
in terms of some as-yet unspecified two-point ``propagator'' $G_{aa'}(x,x')$. This ansatz reduces our problem to the search for a propagator whose properties imply our requirements. We find that very different propagators arise depending on the parity of $d$.

For later convenience, it will be convenient to denote the integral portion of \eqref{Fhat} as being generated, via $F_{ab}^\subS \equiv 2\nabla_{[a} A_{b]}^\subS$, by the vector potential
\begin{equation}
        A_{a}^\subS \equiv \int  G_{aa'}(x,x') J^{a'}(x') dV'.
        \label{FSDef}
\end{equation}
The ``$S$'' here has historically been short for ``singular'' \cite{DetweilerWhiting2003}, as $A_a^\subS$ is indeed singular for pointlike sources, at least when using the Detweiler-Whiting propagator described below. Here, we are not considering point particle sources, so $A_a^\subS$ is not typically singular. It is more appropriate to instead interpret this as a (propagator-dependent) definition for the ``bound portion'' of a body's self-field. It generalizes what is sometimes referred to as the ``Coulomb portion'' of the field.

\subsection{Choosing a propagator}

When $d=4$, well-behaved effective fields are known \cite{HarteEM, HarteReview} to be generated by a certain propagator $G_{aa'}^\subDW$, referred to as the Detweiler-Whiting Green function \cite{DetweilerWhiting2003}. Setting $G_{aa'} = G_{aa'}^\subDW$ in \eqref{Fhat} fixes a precise, physically-reasonable definition for the momentum and its corresponding laws of motion---laws which admit a well-defined point particle limit, well-controlled multipole expansions to all orders for the force and torque, and other desirable properties. Although we discuss Detweiler-Whiting Green functions more explicitly in Section \ref{Sect:even} below, it is convenient at this stage to characterize them implicitly, via three of their properties:
\begin{enumerate}
        \item $G_{aa'}^\subDW (x,x') = 0$ for all timelike-separated $x$, $x'$,

        \item $G_{aa'}^\subDW (x,x') = G_{a'a}^\subDW (x',x)$,

        \item $G_{aa'}^\subDW(x,x')$ is a Green function for the Lorenz-gauge vector potential $A_a(x)$.
\end{enumerate}
Although these are sometimes referred to as the Detweiler-Whiting axioms, they were originally found by Poisson \cite{PoissonLR}. Any propagator which satisfies them induces a field transformation $F_{ab} \mapsto \hat{F}_{ab}$ that can be shown \cite{HarteEM} to imply the identity \eqref{fVary}. There is a precise sense in which they imply laws of motion derivable from \eqref{PDotGenHat}, implying that self-interacting charges act like effective test charges in the field $\hat{F}_{ab}$. Moreover, since $G_{ab}^\subDW$ is a Green function, the associated effective field is source-free in a neighborhood of $\mathfrak{W}$, just like the external Newtonian potential $\hat{\phi}$.

It was noted in \cite{HFTPaper1} that the arguments used to establish these results in four dimensions trivially generalize to any number of dimensions, at least if a propagator satisfying the above axioms does indeed exist. Such a propagator does exist, at least in finite regions, for all \textit{even} $d \geq 4$. However, existence appears to fail when $d$ is odd.

We resolve this by finding an appropriate generalization of the above axioms---valid for all $d \geq 3$, both even and odd---and then constructing explicit propagators which satisfy those axioms. Note that throughout, although we refer to certain statements as axioms, these are to be understood merely as vehicles with which to organize and interpret our results. They are not axioms in the sense of being unproven assumptions. All of our results are derived from first principles.

\subsubsection{Generalizing the axioms}
\label{Sect:genDW}

Of the three axioms stated above, it is the third which is most easily modified. To be more precise, that axiom requires that $G_{aa'}^\subDW$ satisfy
\begin{equation}
        \nabla^b \nabla_b G_{aa'}^\subDW - R_{a}{}^{b} G_{ba'}^\subDW = - \omega_{d-1} g_{aa'} \delta(x,x'),
        \label{Ggreen}
\end{equation}
where $R_{ab}(x)$ denotes the Ricci tensor and $g_{aa'}(x,x')$ the parallel propagator. The differential operator on the left-hand of this equation is motivated by the Maxwell equation
\begin{equation}
        \nabla^b \nabla_b A_a - R_{a}{}^{b} A_{b} = - \omega_{d-1} J_a
\end{equation}
for a Lorenz-gauge vector potential. Demanding that $G_{aa'}^\subDW$ be a Green function in this sense is useful because it may be shown to guarantee that under very general conditions, $\hat{F}_{ab}$ varies slowly inside each cross-section $\mathfrak{B}_\tau$ of a body's worldtube. It therefore ensures that the effective field generated by $G_{aa'}^\subDW$ is not only associated with the law of motion \eqref{PDotGenHat}, but also that generalized forces in that equation admit the well-controlled multipole expansions which are so essential to practical computations.

Multipole expansions such as these can be maintained by supposing that $G_{aa'}$ is not necessarily a Green function, but rather a more general type of parametrix \cite{HFTPaper1}. The right-hand side of \eqref{Ggreen} would then be replaced by
\begin{equation}
        - \omega_{d-1} [ g_{aa'} \delta(x,x') + \mathcal{S}_{aa'}(x,x')],
\end{equation}
where $\mathcal{S}_{aa'}(x,x')$ is sufficiently smooth and satisfies certain other constraints required to maintain the validity of \eqref{fVary}. Such generalizations can be useful because i) parametrices are more easily computed than Green functions, and ii) there may be topological obstructions to constructing Green functions, even when $d=4$. Nevertheless, allowing for a nonzero $\mathcal{S}_{aa'}$ is still not sufficient to solve the odd-dimensional self-force problem; a further generalization is needed.

The generalization we choose is motivated by a desire to demand only what is directly needed, namely that $\hat{F}_{ab}$ ``vary slowly'' throughout each $\mathfrak{B}_\tau$. Although this statement is imprecise as it stands, we note that in the limit that a body's size becomes arbitrarily small, a continuous field cannot vary significantly in any single cross-section. Smoothness in a point-particle limit may thus be used as a proxy for slow variation in more general contexts.

We now replace the three axioms described above by demanding the existence of a propagator $G_{aa'}(x,x')$ with the four properties:
\begin{enumerate}
        \item $G_{aa'} (x,x') = 0$ for all timelike-separated $x$, $x'$.

        \item $G_{aa'} (x,x') = G_{a'a} (x',x)$.

        \item $G_{aa'}(x,x')$ is constructed only from the geometry and depends only quasilocally on the metric, in a sense defined below.

        \item For any point charge moving on a smooth timelike worldline, the source $\omega_{d-1}^{-1} \nabla^b \hat{F}_{ab}$ for the effective field defined by
\eqref{Fhat} is itself smooth, at least in a neighborhood of that worldline.
\end{enumerate}
The first two of these axioms are unchanged from those given by Poisson \cite{PoissonLR}. Axiom 3 is similar to one employed in \cite{HFTPaper1}, while Axiom 4 is new. Axiom 3 demands more precisely that for any vector field $\psi^a$, the Lie derivative $\mathcal{L}_\psi G_{aa'}(x,x')$ can be written as a functional which depends only on the Lie derivative of the metric, and only in a compact region determined by $x$ and $x'$. If considerations are restricted to a single flat spacetime, Axiom 3 may be simplified by demanding simply that $G_{aa'}$ be Poincar\'{e}-invariant.

Physically, Axiom 2 describes a type of reciprocity in the self-field definition associated with $G_{aa'}$ \cite{HarteReview}. It is essential to the establishment of \eqref{fVary}, and thus to the renormalized laws of motion encoded in \eqref{PDotGenHat}. Axioms 1 and 3 guarantee that the renormalizations inherent in those laws of motion involve only physically-acceptable degrees of nonlocality.

As suggested above, our fourth axiom provides a sense in which the renormalized laws of motion can admit well-behaved multipole expansions. It suggests that the $\mathcal{F}^\subEM_\tau[\hat{F}_{ab},J_c; \xi_d]$ appearing in \eqref{PDotGenHat} is generally simpler to evaluate than its bare counterpart $\mathcal{F}^\subEM_\tau[F_{ab},J_c; \xi_d]$. Although Axiom 4 refers only to point particles, these should be interpreted as ``elementary currents'' whose effects can be summed over---as is common in kinetic theory---to yield an overall field for a nonsingular extended charge distribution $J_a$. If the effective field associated with each elementary current is sufficiently smooth, the short-distance behavior associated with any given $J_a$ is considerably suppressed by the appropriate convolution integral. Indeed, there is no obstruction to replacing Axiom 4 by a statement which demands somewhat less regularity. We note as well that there is a sense in which Axiom 4 is ``gauge-agnostic,'' unlike the statement that the Detweiler-Whiting Green function must satisfy the gauge-fixed equation \eqref{Ggreen}.

Now, any $G_{aa'}$ which satisfies our four axioms provides a useful definition for the generalized momentum $\hat{P}_\tau$ associated with an extended body. Moreover, the laws of motion associated with this momentum admit well-behaved multipole expansions. Our next task is to show that such propagators actually exist. It is easily established that any Detweiler-Whiting Green function $G_{aa'}^\subDW$ satisfies our axioms, so that choice can be made whenever such a Green function exists---i.e., when $d$ is even. The freedom to choose other propagators can nevertheless be useful even in those cases. This freedom is however essential when $d$ is odd.

\subsubsection{Even-dimensional propagators}
\label{Sect:even}

If $d \geq 4$ is even, the four axioms given in Section \ref{Sect:genDW} are satisfied by a Green function $G_{aa'} = G_{aa'}^\subDW$ which directly generalizes the four-dimensional Detweiler-Whiting Green function known from \cite{DetweilerWhiting2003, PoissonLR}. These generalizations have the more-explicit form
\begin{equation}
        G_{aa'}^\subDW = \frac{1}{2} \left[ U_{aa'} \delta^{(d/2-2)}( \sigma ) +V_{aa'}\,\Theta(\sigma) \right],
        \label{evenD}
\end{equation}
where $U_{aa'}$ and $V_{aa'}$ are smooth symmetric bitensors which depend only quasilocally on the metric. Essentially the same bitensors also appear in the retarded and advanced Green functions, although there they are to be evaluated only when their arguments are timelike or null-separated. A more direct specification for the bitensors appearing in the Detweiler-Whiting Green function may be found by substituting \eqref{evenD} into \eqref{Ggreen}, which results in the equations collected in Appendix \ref{app:Hadamardeven}.

It is easily shown that if the spacetime is Minkowski, $\nabla^a \nabla_a \sigma = d$. Substituting this into \eqref{DeltaDef}, one finds that the van Vleck determinant is everywhere constant: $\Delta = 1$. Moreover, $\nabla^b \nabla_b (\Delta^{1/2} g_{aa'} ) = 0$, implying that the unique nonsingular solutions to the Hadamard transport equations \eqref{Un} are $\mathsf{U}^{\{ n \} }_{aa'} = 0$ for all $n \geq 1$. Moreover, \eqref{Vtrans} implies that $V_{aa'} = 0$ when its arguments are null-separated. Combining this with \eqref{BoxV}, \eqref{Uexpand}, and \eqref{U0}, it follows that
\begin{equation}
        U_{aa'} = \alpha_d g_{aa'} , \qquad V_{aa'} = 0
        \label{UVflat}
\end{equation}
everywhere in even-dimensional Minkowski spacetimes, where the dimension-dependent constant $\alpha_d$ is explicitly
\begin{equation}
        \alpha_d \equiv \frac{ (-1)^{d/2} 2^{\lambda_d} \sqrt{\pi} }{ \Gamma ( 1/2-\lambda_d) }
        \label{alphaEven}
\end{equation}
in terms of
\begin{equation}
        \lambda_d \equiv 1-d/2.
        \label{lambdaDef}
\end{equation}
Substitution of these results into \eqref{evenD} fully specifies the flat-spacetime, even-dimensional Detweiler-Whiting Green functions. They can also be characterized somewhat differently in this special case, in terms of the advanced and retarded solutions to \eqref{Ggreen}: $G_{aa'}^\subDW = \frac{1}{2} ( G_{aa'}^\ret + G_{aa'}^\adv)$. If $F_{ab}$ is taken to equal the body's retarded field $F^\ret_{ab}$, it follows from \eqref{Fhat} that the effective field $\hat{F}_{ab}$ which determines how bodies move coincides with the so-called radiative field $\frac{1}{2} (F^\ret_{ab} - F_{ab}^\adv)$.

Similar relations between Detweiler-Whiting and advanced and retarded Green functions do not generalize to curved spacetimes, essentially because Huygens' principle is violated; the ``tail'' $V_{aa'}$ is typically nonzero. Although few closed-form results for $U_{aa'}$ and $V_{aa'}$ are known in curved spacetimes, $U_{aa'} = \Delta^{1/2} g_{aa'}$ whenever $d=4$. The bitensor $V_{aa'}$ is also known for $d=4$ plane wave spacetimes \cite{PWtails}, although it is ``pure gauge'' in the sense that $\nabla_{[a} V_{b]b'} = 0$. Expressions in maximally-symmetric spacetimes with arbitrary $d$ may also be extracted from the results of \cite{AllenG}. More generally, numerical or perturbative methods can be used to solve the equations in Appendix \ref{app:Hadamardeven}.

We have already alluded to our four axioms being more general than the original Detweiler-Whiting axioms. This generality is associated with a lack of uniqueness, meaning that other propagators besides \eqref{evenD} are possible when $d$ is even. For example, it is acceptable to choose any propagator with the form
\begin{equation}
        G_{aa'} = G_{aa'}^\subDW + U_{aa'} K(2\sigma/\ell^2) \Theta(\sigma),
        \label{evenDalt}
\end{equation}
where $K$ is some smooth function which vanishes in a neighborhood of zero and $\ell >0$ is a constant lengthscale. If $K$ is fixed, each choice for $\ell$ defines a different propagator, a different $\hat{P}_{\tau}$, a different effective field $\hat{F}_{ab}$, and a different generalized force $\mathcal{F}_\tau$. These differences do not, however, signal any kind of contradiction. Physical consistency is maintained by the fact that \textit{all} of these quantities vary simultaneously, and in very particular ways. Differing forces arise, for example, because they describe rates of change associated with slightly different aspects of the same physical system. One might experimentally associate a particular value of a coupling parameter---such as a mass---with measurements which assume one value of $\ell$, although the same experiments performed on the same system would generically yield a different value when inferred using a different choice of $\ell$; a particular propagator must be fixed before even attempting to interpret experimental data. Nevertheless, there is a sense in which ``true'' observables do not depend on these choices. Further discussion may be found in \cite{HFTPaper1}.

\subsubsection{Odd-dimensional propagators}

If $d \geq 3$ is odd, no Detweiler-Whiting Green function appears to exist. It is thus essential to exploit the freedom afforded by the four axioms listed above. Before constructing an odd-dimensional propagator which satisfies those axioms, note that the retarded Lorenz-gauge Green function in this context has the form
\begin{equation}
        G_{aa'}^\ret = [(-2\sigma)^{\lambda_d} U_{aa'} \Theta (-\sigma)]_\ret ,
        \label{GretOdd}
\end{equation}
where $\lambda_d$ is again given by \eqref{lambdaDef}. The retarded Green function here involves a bitensor $U_{aa'}$ which may be shown to be symmetric and to depend only quasilocally on the metric. Also note that the ``$\ret$'' on the whole expression denotes that it has support only for $x'$ in the past of $x$. As in the even-dimensional context, $U_{aa'} = \alpha_d g_{aa'}$ in Minkowski spacetime, although the odd-dimensional constants here are given by
\begin{equation}
        \alpha_d \equiv \frac{(-1)^{1/2+ \lambda_d} \Gamma(-\lambda_d)}{ \sqrt{\pi} \Gamma(1/2-\lambda_d)  }
        \label{alphaOdd}
\end{equation}
instead of \eqref{alphaEven}. In more general spacetimes, a prescription to compute $U_{aa'}$ is described in Appendix \ref{app:Hadamardodd}.

Whether in Minkowski spacetime or not, it is evident from \eqref{GretOdd} that Huygens' principle is violated when $d$ is odd. Signals travel not only on null cones, but also inside of them. Although Huygens' principle is similarly violated for Maxwell fields in curved even-dimensional spacetimes, the odd-dimensional case is different in that $G_{aa'}^\mathrm{ret}$ is unbounded even when its arguments are timelike-separated. Indeed, the tail here is not even locally integrable in general. Eq. \eqref{GretOdd} is thus closer to a schematic than a precise description for the retarded Green function. The correct distributional solution can more precisely be constructed by considering $[(-2\sigma)^\lambda U_{aa'} \Theta(-\sigma)]_\ret$ for values of $\lambda$ in which the singularity is integrable and then analytically continuing the result to $\lambda \rightarrow \lambda_d$. Another mathematical detail is that $[(-2\sigma)^{\lambda} \Theta(-\sigma)]_\ret$ should be regarded as a single symbol, not a product of singular distributions. These and other details associated with the odd-dimensional retarded Green functions are made precise in, e.g., \cite{FriedlanderWave, GelfandShilov}.

The propagators which allow us to solve the self-force problem in odd numbers of dimensions can also defined using analytic continuation. They are
\begin{align}
        G^\mathrm{odd}_{aa'} \equiv  \frac{ (-1)^{\frac{1}{2}-\lambda_d} }{ 2\pi }  U_{aa'}  \lim_{\lambda \rightarrow \lambda_d}  \ell^{2\lambda}\frac{\partial}{\partial \lambda} \left[ ( 2 \sigma / \ell^2 )^\lambda \Theta (\sigma)\right],
        \label{EqLogProp0}
\end{align}
where the overall prefactor has been chosen in order to enforce Axiom 4. The $U_{aa'}$ appearing here is constructed in the same way as for the retarded and advanced Green functions. Performing the differentiation in \eqref{EqLogProp0} while leaving the limit $\lambda \to \lambda_d$ implicit, our propagator can alternatively be written as
\begin{align}
        G^\mathrm{odd}_{aa'} = \frac{ (-1)^{ \frac{1}{2} (d-1) } }{2\pi} (2 \sigma)^{\lambda_d} U_{aa'} \ln (2\sigma/\ell^2) \Theta(\sigma).
                \label{EqLogProp}
\end{align}
In either form, these expressions fix a 1-parameter family of propagators which depend on an arbitrary lengthscale $\ell > 0$. This lengthscale is introduced in order to ensure that the quantity differentiated with respect to $\lambda$ is dimensionless in \eqref{EqLogProp0}. Choosing different values for $\ell$ would result in propagators which differ by multiples of $U_{aa'} (2\sigma)^{\lambda_d} \Theta(\sigma)$, a propagator which generates source-free solution to Maxwell's equations. Although variations in $\ell$ generically change effective fields and thus forces, such shifts have no observable consequences. They merely parametrize different ways to describe the same physical system. The situation here is fully analogous to that associated with the $\ell$-dependence of \eqref{evenDalt} and also with the non-uniqueness of the static propagators discussed in \cite{HFTPaper1}.

We now verify that the propagator $G_{aa'}^\mathrm{odd}$ satisfies the four axioms described in Section \ref{Sect:genDW}. That the first of these holds is immediately clear from the presence of the $\Theta$-function in \eqref{EqLogProp}. The second and third axioms are verified by noting that $\sigma$ and $U_{aa'}$ are symmetric in their arguments and depend quasilocally on the metric, as elaborated in Appendix \ref{app:Hadamard}.

Considerably more effort is required to show that our propagator also satisfies Axiom 4. We do so using the direct calculations summarized in Appendix \ref{app:ppOdd}: Consider a point particle with timelike worldline $\Gamma$ and let $F_{ab}$ be identified with that particle's retarded field. Then the $\hat{F}_{ab}$ generated by $G_{aa'}^\mathrm{odd}$ is given by combining \eqref{AhatT} and \eqref{FhatT}. In those equations, the only position dependence is via smooth functions, at least if the metric and the worldline are themselves smooth. We thus conclude that $\hat{F}_{ab}$, and hence its source $\omega_{d-1}^{-1} \nabla^b \hat{F}_{ab}$, must be smooth in the presence of retarded boundary conditions.  Repeating the problem with more general boundary or initial conditions would merely change $\hat{F}_{ab}$ by a homogeneous solution to Maxwell's equations. The source is thus smooth in general, verifying Axiom 4. As claimed, all axioms given in Section \ref{Sect:genDW} are satisfied by the propagator \eqref{EqLogProp0}.

Although $G_{aa'}^\mathrm{odd}$ is not a Green function or more general parametrix for Lorenz-gauge vector potentials, some intuition for it may nevertheless be gained by noting that the derivative with respect to $\lambda$ which appears in its definition \eqref{EqLogProp} evinces a procedure which ``infinitesimally varies $d$.'' This suggests that our map $F_{ab} \mapsto \hat{F}_{ab}$ may reduce to dimensional regularization in a point particle limit, and may provide an underlying physical and mathematical origin for that procedure at least in the present context. We are not aware of any other examples in which dimensional regularization arises as the natural limit of a more-general nonsingular operation which follows from first principles.

\subsection{Laws of motion}
\label{Sect:LawsofMotion}

To summarize our development at this point, we have shown that for all $d \geq 3$, two-point propagators $G_{aa'}$ may be found which satisfy the four axioms given in Section \ref{Sect:genDW}. If $d$ is even, one possibility is to set $G_{aa'} = G_{aa'}^\subDW$, where $G_{aa'}^\subDW$ is given by \eqref{evenD}. If $d$ is odd, one may instead use $G_{aa'} = G_{aa'}^\mathrm{odd}$, where $G_{aa'}^\mathrm{odd}$ satisfies \eqref{EqLogProp0}. Regardless, any specific choice for $G_{aa'}$ which satisfies the given axioms may be associated with a particular renormalization $\hat{P}_\tau$ of the bare generalized momentum defined by \eqref{PDef}. More specifically, the methods reviewed in \cite{HarteReview} may be used to show that the appropriate relation between these momenta is
\begin{widetext}
\begin{align}
        \hat{P}_\tau = P_\tau + \frac{1}{2} \left( \int_{\mathfrak{B}_\tau^+} \! \!  dV J^a \mathcal{L}_\xi \int_{\mathfrak{B}_\tau^-} \! \!  dV' G_{aa'} J^{a'} - \int_{\mathfrak{B}_\tau^-} \! \!  dV J^a \mathcal{L}_\xi \int_{\mathfrak{B}_\tau^+} \! \!  dV' G_{aa'} J^{a'} \right) + \int_{\mathfrak{B}_\tau} \! \! dS_a J^a
        \nonumber
        \\
        ~ \times \left(\int_{\mathfrak{B}_\tau} \! \! dV' \xi^b G_{bb'} J^{a'} - \int_0^1 du u^{-1} \nabla_{b'} \sigma F^{b'c'}_\subS \xi_{c'} \right),
        \label{Prenorm}
\end{align}
\end{widetext}
where $\mathfrak{B}_\tau^\pm$ denotes the portion of the body's worldtube which lies to the future ($+$) or past ($-$) of $\mathfrak{B}_\tau$, and the primes in the $u$-integral are associated with points on the same geodesic $y'(u)$ which appeared in \eqref{PDef}. The important point here is that the renormalizing terms are appropriately-local: The first of our axioms for $G_{aa'}$ implies that the momentum at time $\tau$ can depend on $T^{ab}_\subB$, $J_a$, and $F_{ab}$ only in the body's worldtube, and only on those portions of the worldtube which are spacelike or null-separated from $\mathfrak{B}_\tau$. This is in strong contrast to any attempt which might be made to directly compute a ``self-momentum'' associated with $T^{ab}_\subEM$. Nevertheless, the two procedures do coincide in simple cases where nonlocality is not an issue; see \cite{HarteEM} for the $d=4$ discussion.

Continuing our summary, fixing an appropriate $G_{aa'}$ fixes a particular definition for $\hat{P}_\tau$, and we have shown that this momentum must satisfy the laws of motion \eqref{PDotGenHat}. These laws are instantaneously identical to those which hold for an extended test body with stress-energy tensor $\hat{T}^{ab}_\subB$ and current density $J_a$, coupled to a spacetime metric $g_{ab}$ and an electromagnetic field $\hat{F}_{ab}$. The effective electromagnetic field here depends on $G_{aa'}$ and is given more precisely by \eqref{Fhat}. The renormalized stress-energy $\hat{T}^{ab}_\subB$ also depends on $G_{aa'}$, and at least in static contexts, it can be written in terms of functional derivatives of the appropriate propagator \cite{HFTPaper1}.

Regardless, once a propagator has been fixed, the laws of motion are fixed as well. The force on a body may be found by computing $\hat{F}_{ab}$ from $F_{ab}$ and substituting the result into an appropriate test body equation. For example, the lowest-order electromagnetic force acting on a body with charge $q$ is given by the usual Lorentz expression
\begin{equation}
        \hat{f}_a = q \hat{F}_{ab} \dot{\gamma}^b .
        \label{fLorentz}
\end{equation}
Similarly, the lowest-order electromagnetic torque on a body with electromagnetic dipole moment $q^{ab} = q^{[ab]}$ is
\begin{equation}
        \hat{n}^{ab} = 2 q^{c[a} \hat{F}^{b]}{}_{c} .
        \label{nDipole}
\end{equation}
Although these expressions might appear superficially similar to test-body expressions, they encode all leading-order self-force and self-torque effects in general spacetimes.

More generally, the full multipole expansion for the electromagnetic generalized force can be shown to be
\begin{align}
        \mathcal{F}^\subEM_\tau [ \hat{F}_{ab} &, J_c; \xi_d] = q \hat{F}_{ab} \xi^a \dot{\gamma}^b
 \nonumber
 \\
 & ~ + \sum_{n=1}^\infty \frac{n}{(n+1)!} q^{b_1 \cdots b_n a} \mathcal{L}_\xi \hat{F}_{ab_1,b_2 \cdots b_n} ,
 \label{EMforce}
\end{align}
where $q^{b_1 \cdots b_n a}$ denotes the $2^n$-pole moment of $J^a$ and $\hat{F}_{ab,c_1 \cdots c_n}$ the $n$th tensor extension of $F_{ab}$. Letting $\hat{I}^{c_1 \cdots c_n ab}$ denote the $2^n$-pole moment of $\hat{T}^{ab}_\subB$ and $g_{ab,c_1 \cdots c_n}$ the $n$th tensor extension of $g_{ab}$, the gravitational generalized force may be similarly expanded as
\begin{align}
                \mathcal{F}^\subG_\tau[\hat{T}^{ab}_\subB;\xi_c] = \frac{1}{2} \sum_{n=2}^\infty \frac{1}{ n!} \hat{I}^{c_1 \cdots c_n ab} \mathcal{L}_\xi g_{ab,c_1 \cdots c_n} .
                \label{Gravforce}
\end{align}
Tensor extensions are discussed in more detail in \cite{Dix74, HarteReview}; the first nontrivial ones are
\begin{equation}
        g_{ab,cd} = \frac{2}{3} R_{a(cd)b}, \qquad F_{ab,c} = \nabla_c F_{ab}.
\end{equation}
Regardless, the gravitational expression here involves only quadrupole and higher moments, and the the tangent $\dot{\gamma}^a$ to the reference worldline appears explicitly only in the Lorentz force (and not in the higher-order electromagnetic terms or in any gravitational terms).

Eqs. \eqref{EMforce} and \eqref{Gravforce} may now be combined with \eqref{PDotGenHat} to yield the full laws of motion. It is however more conventional to split $\hat{P}_\tau$ into its linear and angular components via a ``hatted'' analog of \eqref{pSDef}. Doing so, it is convenient to define a renormalized force $\hat{f}_a$ and a renormalized torque $\hat{n}_{ab} = \hat{n}_{[ab]}$ using the similar implicit equation
\begin{equation}
        \frac{d}{d\tau} \hat{P}_\tau[\xi_a]= \hat{f}_a \xi^a + \frac{1}{2} \hat{n}^{ab} \nabla_a \xi_b.
\end{equation}
This definition provides forces and torques which measure the degree by which the Mathisson-Papapetrou equations are violated:
\begin{gather}
        \frac{D}{d\tau} \hat{p}^a = \frac{1}{2} R_{bcd}{}^{a} \hat{S}^{bc} \dot{\gamma}^d + \hat{f}^a,
        \label{pDot}
        \\
        \frac{D}{d\tau} \hat{S}^{ab} = 2 \hat{p}^{[a} \dot{\gamma}^{b]} + \hat{n}^{ab}.
        \label{sDot}
\end{gather}
That the first term on the right-hand side of the second equation is not considered a torque is natural in the sense that an analogous term exists even for the angular momentum of an isolated system in Newtonian physics. This is so essentially because a Euclidean rotation about one origin can be decomposed into a rotation about another origin plus a translation. If the origin about which the angular momentum is defined is moving, it must ``mix'' over time with the linear momentum conjugate to the translations generated by that motion. The $\frac{1}{2} R_{bcd}{}^{a} \hat{S}^{bc} \dot{\gamma}^d$ term on the right-hand side of \eqref{pDot} is similarly interpreted as arising from the fact that in a curved spacetime, pure translations at one point are not necessarily pure translations at another point. Both this term and the $2 \hat{p}^{[a} \dot{\gamma}^{b]}$ in \eqref{sDot} are thus kinematic in origin, an interpretation which persists even in the absence of any true symmetries.

Now, explicit multipole expansions for our force and torque may be derived by combining \eqref{PDotGenHat}, \eqref{EMforce}, \eqref{Gravforce}, \eqref{pDot}, and \eqref{sDot} while varying over all GKFs $\xi_a$. The result is no different than it is when $d=4$, and is given by Eqs. (193) and (194) of \cite{HarteReview}. The monopole truncation for the resulting force is simply \eqref{fLorentz}, while the dipole truncation for the torque is \eqref{nDipole}. Gravitational effects do not enter until quadrupole order. To all multipole orders, our expansions for $\hat{f}_a$ and $\hat{n}_{ab}$ are structurally identical to the multipole expansions derived by Dixon for an extended test body \cite{Dix74}. All differences are implicit in our hat notation, which alters the definitions for the momenta, the stress-energy moments, and the electromagnetic field in such a way that multipole expansions remain useful even in the presence of strong self-interaction.

Thus far, all of our discussion has allowed for essentially-arbitrary reference worldlines $\gamma(\tau)$ and foliating hypersurfaces $\mathfrak{B}_\tau$. It is however conventional to identify the worldline with some kind of mass center and the foliation with the instantaneous rest frames associated with that center. The first of these demands is typically accomplished by imposing a ``spin supplementary condition'' which asks that the mass dipole moment associated with the body vanish in an appropriate reference frame. There are different ways to make this precise. Although it is not essential, here we do so by choosing $\gamma(\tau)$ such that
\begin{equation}
        \hat{S}^{ab} \hat{p}_b = 0.
        \label{CM}
\end{equation}
We can also fix the foliation by demanding that each $\mathfrak{B}_\tau$ is constructed from the hyperplane formed by all geodesics which pass through $\gamma(\tau)$ and are orthogonal to $\hat{p}^a(\tau)$ at that point. These conditions may now be used to relate $\hat{p}^a$ to $\dot{\gamma}^a$; they are not necessarily parallel. Differentiating \eqref{CM} while using \eqref{pDot} and \eqref{sDot}, the momentum-velocity relation is found to be
\begin{equation}
        \hat{m} \dot{\gamma}^a  = \frac{1}{\hat{m}} (\mathcal{I}^{-1})^{a}{}_{b} [ (-\hat{p} \cdot \dot{\gamma}) \hat{p}^b - \hat{S}^{bc} \hat{f}_c^\circ - \hat{n}^{bc} \hat{p}_c ],
        \label{momVel}
\end{equation}
where we have defined the renormalized mass by
\begin{equation}
        \hat{m}^2 \equiv - \hat{p}^a \hat{p}_a,
        \label{mDef}
\end{equation}
used the inverse of
\begin{equation}
        \mathcal{I}^{a}{}_{b} \equiv \delta^a_b + \frac{1}{\hat{m}^2} \hat{S}^{ac} \left( \frac{1}{2}  R_{bcdf} \hat{S}^{df} - q \hat{F}_{bc} \right),
        \label{IDef}
\end{equation}
and let $\hat{f}^\circ_a \equiv \hat{f}_a - q \hat{F}_{ab} \dot{\gamma}^b$ be the non-Lorentz portion of the force (which is relevant because the Lorentz force is the only component which depends explicitly on $\dot{\gamma}^a$). A more explicit momentum-velocity relation can be obtained if the matrix rank of $\hat{S}_{ab}$ is no greater than two \cite{EhlersRudolph, HarteReview}, although such a condition can be guaranteed only when $d < 5$. Eq. \eqref{momVel} may instead be applied whenever $\mathcal{I}^{a}{}_{b}$ is invertible. Components of $\hat{p}^a$ which fail to be parallel to $\dot{\gamma}^a$ are referred to as hidden momentum \cite{Bobbing, CostaReview}. Although the equations presented here are complicated, they differ from their test-body counterparts only via physically-ignorable renormalizations and the nonlocal map $F_{ab} \mapsto \hat{F}_{ab}$. No simpler result could reasonably be expected, at least in the absence of a particular approximation scheme.

\section{Point particles in flat spacetimes}
\label{Sect:Flat}

One useful class of approximations may be interpreted as point particle limits. Certain limits of this type have been discussed in detail in \cite{GrallaHarteWald} when $d=4$, while others, valid for all $d$, were considered in \cite{HFTPaper1}. Regardless of details, one considers a 1-parameter family of extended bodies whose sizes scale linearly with a control parameter $\delta >0$ which is eventually taken to zero. Various physical constraints require that other properties of the bodies---such as their net charges---scale at rates which depend on particular powers of $\delta$, powers which depend both on $d$ and on the specific property being considered. Reasonable motivations can be found for different approximations, although a general feature is that self-force effects can be ``more important'' in lower numbers of dimensions; they generically compete in magnitude with test-body effects associated with lower multipole orders. Conversely, leading-order self-force effects are strongly suppressed relative to leading-order test-body effects when $d$ is large. Self-interaction should thus be understood not in isolation, but in combination with test-body effects up to an appropriate multipole order. Nevertheless, our discussion below focuses for simplicity mainly on the computation of leading-order self-forces and self-torques.

We now apply the results derived in Section \ref{Sect:NonPert} to perform these computations for ``point particles'' in Minkowski spacetimes of various dimensions. Although we have in mind a point particle limit, we do not discuss details of the associated family of extended charges. Instead, we suppose that in this limit, the family of worldtubes associated with the extended bodies used to construct the point particle limit shrink to a timelike worldline $\Gamma = \{ \gamma(\tau) : \tau \in \mathbb{R} \}$, where the parametrization has been chosen such that $\dot{\gamma}^a \dot{\gamma}_a = -1$. We take $\Gamma$ to be the reference worldline for the constructions of the previous section, and assume that it satisfies the spin supplementary condition \eqref{CM}. In the limit, the bodies' net charges typically tend to zero along with their diameters; a body with too much charge for its size and mass cannot hold itself together without exerting stresses which violate energy conditions. Regardless, it is convenient to consider a point particle limit in which the current densities associated with members of the given family of extended charge distributions approach an appropriate function of $\delta$ multiplied by the point-particle current density
\begin{equation}
        J_\mathrm{pp}^a(x) = q \int \dot{\gamma}^a(\tau) \delta(x,\gamma(\tau)) d\tau .
        \label{Jpoint}
\end{equation}
The $q$ appearing here is a fixed parameter which represents a $\delta$-dependent rescaling of the charges associated with different members of the family in the limit $\delta \to 0^+$. Despite this, we refer to it as ``the'' charge for simplicity. Leading-order self-forces and self-torques may now be computed by evaluating the effective field $\hat{F}_{ab}$ associated with $J^a_\mathrm{pp}$ and then inserting the result into \eqref{fLorentz} and \eqref{nDipole}. No regularization is required.

\subsection{Even dimensions}
\label{Sect:forceEven}

In even-dimensional Minkowski spacetimes, the prescription described in Section \ref{Sect:NonPert} implies that it is useful to define a body's renormalized momentum using the propagator $G_{aa'}= G_{aa'}^\subDW$, where $G^\subDW_{aa'}$ is given by \eqref{evenD}. The $S$-field generated by this propagator and associated with a current of the form \eqref{Jpoint} is found by substituting \eqref{UVflat} into \eqref{eq:SingFieldEven}, which yields
\begin{equation}
        A_a^\subS = \frac{q \alpha_d }{2} \sum_{\tau \in \{ \tau_\pm \} } \frac{1}{|\dot{\sigma}|} \left( - \frac{\partial}{\partial \tau} \frac{1}{\dot{\sigma}} \right)^{d/2-2} g_{aa'} \dot{\gamma}^{a'},
        \label{ASeven}
\end{equation}
where the advanced and retarded times $\tau_\pm(x)$ are defined in Appendix \ref{app:CoincidenceLimits} and $\alpha_d$ depends on the dimension via \eqref{alphaEven}. In the special case where the physical field $F_{ab}$ coincides with the particle's retarded field, a vector potential for the effective field $\hat{F}_{ab}$ can be written as in \eqref{eq:HatFieldEven}. Specializing that equation to flat spacetime,
\begin{equation}
\label{AhatEvenFlat}
        \hat{A}_a = \left. \frac{q \alpha_d}{2|\dot{\sigma}|} \left( -\frac{\partial}{\partial \tau}\frac{1}{\dot{\sigma}} \right)^{d/2-2} g_{aa'}\dot{\gamma}^{a'} \right|^{\tau = \tau_-}_{\tau = \tau_+} .
\end{equation}
Leading-order self-forces and self-torques may now be computed by evaluating $\hat{F}_{ab} = 2 \nabla_{[a} \hat{A}_{b]}$ on the particle's worldline and substituting the result into \eqref{fLorentz} and \eqref{nDipole}.

If $d=4$, this procedure is reasonably straightforward using the expansion techniques and limits collected in Appendix \ref{app:CoincidenceLimits}; carrying out the relevant calculations results in
\begin{equation}
        \hat{F}_{ab} = \frac{4}{3} q \dot{\gamma}_{[a} \dddot{\gamma}_{b]}
        \label{F4D}
\end{equation}
on the particle's worldline. It follows that the leading-order self-force with retarded boundary conditions in four-dimensional Minkowski spacetime is
\begin{equation}
        \hat{f}_a = \frac{2}{3} q^2 h_{ab} \dddot{\gamma}^b,
        \label{f4D}
\end{equation}
where $h_{ab} \equiv g_{ab} + \dot{\gamma}_a \dot{\gamma}_b$ denotes a projection operator associated with the particle's rest frame. This may be recognized as the standard Abraham-Lorentz-Dirac radiation-reaction force; see, e.g., \cite{Dirac, PoissonLR}. The leading-order four-dimensional self-torque follows immediately as well:
\begin{equation}
        \hat{n}^{ab} = \frac{4}{3} q q^{c[a} (\dot{\gamma}^{b]} \dddot{\gamma}_c -  \dddot{\gamma}^{b]} \dot{\gamma}_c ).
        \label{n4D}
\end{equation}
Although the fields for our particle have been obtained without a dipole moment $q^{ab}$, including one would still result in this self-torque at leading order. Also note that $\hat{n}^{ab} \dot{\gamma}_b$ need not vanish in \eqref{n4D}. Such components may be seen from \eqref{momVel} to induce a hidden momentum in which the direction of $\hat{p}_a$ differs from that of $\dot{\gamma}_a$.

Deriving analogous results in higher numbers of even dimensions is straightforward but tedious. For $d=6$, we find by expanding \eqref{AhatEvenFlat} that the effective field on a particle's worldline is
\begin{align}
        \hat{F}_{ab} = \frac{2}{9} q \bigg( \frac{4}{5} \gamma^{(5)}_{[a} \dot{\gamma}_{b]} + \gamma^{(4)}_{[a} \ddot{\gamma}_{b]} - 2 |\ddot{\gamma}|^2 \dddot{\gamma}_{[a} \dot{\gamma}_{b]}
        \nonumber
        \\
        ~ - \frac{3}{2} \frac{d |\ddot{\gamma}|^2 }{ d\tau} \ddot{\gamma}_{[a} \dot{\gamma}_{b]} \bigg),
        \label{F6D}
\end{align}
implying that the leading-order flat-spacetime self-force with retarded boundary conditions is
\begin{align}
        \hat{f}^a =-\frac{1}{9}q^{2}h^{ab}\left( \frac{4}{5}\gamma_{b}^{(5)}-2|\ddot{\gamma}|^{2}\dddot{\gamma}_{b}-3 \frac{d|\ddot{\gamma}|^2}{d\tau} \ddot{\gamma}_{b}\right).
\end{align}
This force agrees with expressions which have been obtained elsewhere using different methods \cite{GaltsovSpirin, Kazinski, Kosyakov6D, Galakhov, Ofek1}. Our approach trivially allows a self-torque to be obtained as well, by substituting \eqref{F6D} into \eqref{nDipole}, although we omit this for brevity.

Continuing, the $d=8$ effective field with retarded boundary conditions may be computed by again expanding \eqref{AhatEvenFlat} using the techniques of Appendix \ref{app:CoincidenceLimits}. We omit the full result, noting only that the leading-order self-force is
\begin{align}
        \label{eq:forceD8}
        \hat{f}^a = \frac{ 2q^2 }{ 525 }   h^{ab} \bigg[ \!  \gamma^{(7)}_b - 7 |\ddot{\gamma}|^2 \gamma^{(5)}_b - \frac{35}{2} \frac{ d|\ddot{\gamma}|^2}{d\tau} \gamma^{(4)}_b
        \nonumber
        \\
        ~+ \frac{7}{9} \left(25 |\ddot{\gamma}|^4 + 7 |\dddot{\gamma}|^2 - 24 \frac{d^2 |\ddot{\gamma}|^2}{d\tau^2}\right) \dddot{\gamma}_b
        \nonumber
        \\
        ~ + \frac{7}{6} \frac{d}{d\tau} \left( 25 |\ddot{\gamma}|^4 + 7 |\dddot{\gamma} |^2 - 9 \frac{d^2|\ddot{\gamma}|^2}{d\tau^2} \right) \ddot{\gamma}_b   \bigg].
\end{align}
Taking into account differing sign conventions and a typographical error in which $\dot{u}^2 \ddot{u}$ should really be $\ddot{u}^2 \ddot{u}$, this agrees with an expression found in \cite{Galakhov}.

Although our flat-spacetime self-forces agree with existing expressions in Minkowski spacetimes with even numbers of dimensions, our odd-dimensional predictions do not.

\subsection{Odd dimensions}
\label{Sect:forceOdd}

Self-forces and self-torques acting on point charges in odd-dimensional Minkowski spacetimes may now be obtained by fixing the definition for the renormalized momentum by identifying the propagator $G_{aa'}$ with the $G^{\mathrm{odd}}_{aa'}$ given by \eqref{EqLogProp}. The constant lengthscale $\ell$ which appears in the definition for $G^{\mathrm{odd}}_{aa'}$ is assumed to have been fixed as well, although its precise value is irrelevant. With these choices, it is shown in Appendix \ref{app:OddD} that the $S$-field near the worldline of a point charge with current density \eqref{Jpoint} may be expanded in powers of the radar distance $r$ associated with $\Gamma$:
\begin{widetext}
\begin{align}
        A^\subS_a = \frac{\Gamma(\tfrac{d}{2}-1)}{2 \,\pi^{3/2} \Gamma(\tfrac{d}{2}-\tfrac{1}{2})  }  \Bigg\{ \sum_{ n=\frac{1}{2} (d-3) }^\infty \frac{ \Gamma(n+\frac{1}{2}) \Gamma(2-\frac{d}{2}) }{ (2n)! \Gamma(n + \frac{1}{2} (5-d) ) } \Big[\left( H_{n-\frac{1}{2}(d-3)}  - H_{1-\frac{d}{2}} - 2 \ln (r/\ell) \right) W_a^{ \{ 2n \} }  - \partial_\lambda W_a^{ \{ 2n \} }\Big]
        \nonumber
        \\
        ~ \times r^{2n-(d-3)}  + (-1)^{\frac{1}{2}(d-3)} \sum_{n=0}^{\frac{1}{2}(d-5)} \frac{ (-1)^n  \Gamma(n+\tfrac{1}{2}) \Gamma(2-\tfrac{d}{2}) \Gamma(\tfrac{1}{2}(d-3)-n)  }{ (2n)!}  \frac{ W_a^{ \{ 2n \} } }{r^{(d-3)-2n} } \Bigg\}.
        \label{ASfin}
\end{align}
\end{widetext}
Here, the sum in the second line is understood to exist only for $d \geq 5$, $r(x)$ is defined more precisely by \eqref{rDef}, $H_\mu$ denotes the $\mu$th harmonic number, and the coefficients $W_a^{\{n\}}(x;\lambda)$ are defined by \eqref{WcoeffDef} in terms of the flat-spacetime specialization
\begin{equation}
        W_a(x,\tau;\lambda) = q g_{aa'}(x,\gamma(\tau)) \dot{\gamma}^{a'} (\tau) \Sigma^\lambda(x,\tau)
        \label{WDefFlat}
\end{equation}
of \eqref{WDef} and the ``factorized world function'' $\Sigma(x,\tau)$ defined by \eqref{sigExpand}. All implicit instances of $\lambda$ in \eqref{ASfin} are to be evaluated at $\lambda = \lambda_d = 1-d/2$. Some results for the first few $W_a^{\{n\}}$ and their exterior derivatives on the particle's worldline are collected in \eqref{W} and \eqref{DW}. Also note that although we are focusing here on flat spacetimes, the derivation in Appendix \ref{app:Retfield} shows that \eqref{ASfin} is actually valid in all odd-dimensional spacetimes, as long as \eqref{WDefFlat} is replaced by the more-general \eqref{WDef}.

Regardless, \eqref{ASfin} is the odd-dimensional analog of \eqref{ASeven}. It generically involves non-negative even powers of $r$, non-negative even powers multiplied by $\ln r$, and negative even powers down to $r^{-(d-3)}$. The self-force and self-torque may be evaluated by subtracting this from a physical vector potential and then using \eqref{Fhat} to compute $\hat{F}_{ab}$. The result is automatically finite, at least in the absence of impulsive incoming waves or other singular phenomena external to the body itself.

As in the even-dimensional context, it is interesting to suppose that the true electromagnetic field $F_{ab}$ is equal to the retarded field $F_{ab}^\ret$. Assuming this, the relevant subtraction with $F_{ab}^\subS$ is performed in Appendix \ref{app:Hatfield}, which culminates in the effective field \eqref{FhatGenFin}. That result is valid for general odd-dimensional spacetimes. Specializing it to the flat case by introducing Minkowski coordinates $x^\mu$ while using \eqref{GretOdd} and \eqref{alphaOdd}, we find that
\begin{widetext}
\begin{align}
        \hat{F}_{\mu\nu} = \frac{ 2 (-1)^{\frac{1}{2} (d-3)} \Gamma(d/2-1)}{ \sqrt{\pi} \Gamma(\frac{1}{2}(d-1))  }
 \Bigg[ (d-2) q \int^{\tau-\epsilon}_{-\infty} \frac{ X_{[\mu}(\tau,\tau') \dot{\gamma}_{\nu]} (\tau')  }{ [-X^2(\tau,\tau')]^{d/2} } d\tau' - \sum_{n=0}^{d-4} \frac{ (-1)^n }{ n!  } \bigg( \frac{ \nabla_{[\mu} W_{\nu]}^{ \{ n \} } }{ d-3-n } + \frac{1}{\epsilon } \dot{\gamma}_{[\mu} W_{\nu]}^{ \{ n \} } \bigg)
        \nonumber
        \\
        ~ \times \frac{1}{\epsilon^{d-3-n}} - \frac{1}{(d-3)!} \left( \frac{1}{\epsilon} \dot{\gamma}_{[\mu} W_{\nu]}^{ \{ d-3 \} }  - \nabla_{[\mu} W_{\nu]}^{ \{ d-3 \} } \ln (\epsilon/\ell) - \frac{1}{2} \partial_\lambda \nabla_{[\mu} W_{\nu]}^{ \{ d-3 \} } - \frac{1}{(d-2)} \dot{\gamma}_{[\mu} W_{\nu]}^{ \{ d-2 \} } \right) \Bigg]
        \label{FhatFin}
\end{align}
\end{widetext}
on $\Gamma$, where we have omitted an implicit limit $\epsilon \to 0^+$ and defined $X_\mu(\tau,\tau') \equiv \gamma_\mu(\tau) - \gamma_\mu(\tau')$. Although individual terms here involve negative powers of $\epsilon$ and also $\ln \epsilon$, these cancel similarly-divergent terms in the integral; the overall limit here is well-behaved. Also nmote that even though $G_{aa'}^\mathrm{odd}$ is not a Green function and the effective field here is not in general a solution to the source-free Maxwell equations, it is source-free for inertially-moving particles. Indeed, it vanishes in those cases.

Two qualitative differences may now be observed between our flat-spacetime effective fields in even and odd numbers of dimensions. First, the odd-dimensional $\hat{F}_{ab}$ depends on the particle's past history. Its even-dimensional counterpart does not. Second, our odd-dimensional field depends on the arbitrary parameter $\ell > 0$ which appears in the definition for $G_{aa'}^\mathrm{odd}$. Varying $\ell$ results in different propagators, different definitions for a body's momentum, and different forces. In practice, one can choose a convenient value and then use it to infer masses and other parameters from available experimental data. Although those inferences would differ somewhat with different choices for $\ell$, they would do so in predictable ways which could be computed from the expressions found in Section \ref{Sect:NonPert}.

Having now noted that the even and odd-dimensional effective fields discussed here differ both in their history dependence and their parameter dependence, we emphasize that neither of these differences are essential. Parameter dependence can appear for even $d$ if, e.g., one constructs momenta using a family of propagators with the form \eqref{evenDalt}. Furthermore, history dependence generically occurs in even-dimensional effective fields whenever the spacetime is curved. Indeed, it arises even in flat even-dimensional spacetimes if a body is coupled to a massive field (as opposed to the massless Maxwell couplings considered here).

\subsubsection{Special cases}

In the absence of closed-form expressions for the coefficients $W_\mu^{\{n\}}$ and $\nabla_{[\mu} W^{\{n\}}_{\nu]}$ which appear in \eqref{FhatFin}, it is not possible to provide fully-explicit formulae for all odd-dimensional self-forces. However, those coefficients can be computed, for each $n$, using the methods of Appendix \ref{app:CoincidenceLimits}. Explicit self-forces may thus be obtained for any \textit{specific} odd $d$. We now discuss three and five-dimensional Minkowski spacetimes as special cases.

Assuming retarded boundary conditions, substitution of \eqref{W} and \eqref{DW} into \eqref{FhatFin} results in the $d=3$ effective field
\begin{align}
        \hat{F}_{\mu\nu} = 2 q \left[ \int_{-\infty}^{\tau-\epsilon} \!\! \left( \frac{ X_{[\mu} \dot{\gamma}'_{\nu]} }{ (-X^2 )^{3/2} } \right) d\tau'
 + \frac{1}{2} \ddot{\gamma}_{[\mu}\dot{\gamma}_{\nu]}  \ln (\epsilon/e \ell) \right]
        \label{F3D}
\end{align}
on $\Gamma$, where $e$ denotes the base of the natural logarithm and the limit $\epsilon \to 0^+$ has again been left implicit. Combining this with \eqref{fLorentz} immediately yields the leading-order three-dimensional self-force
\begin{align}
        \label{eq:Lorentz3D}
        \hat{f}_\mu = 2 q^2 \left[ \int^{\tau-\epsilon}_{-\infty} \!\! \left(\frac{  X_{[\mu} \dot{\gamma}_{\nu]}' \dot{\gamma}^\nu }{ (-X^2)^{3/2} } \right)d\tau' - \frac{1}{4} \ln(\epsilon/e \ell) \ddot{\gamma}_\mu \right].
\end{align}
Similarly, substituting \eqref{F3D} into \eqref{nDipole} yields the leading-order three-dimensional self-torque
\begin{align}
        \hat{n}^{\mu\nu} = 2 q q^{\rho [ \mu } \Bigg[ \int_{-\infty}^{\tau-\epsilon} \left( \frac{ X^{\nu]} \dot{\gamma}'_{\rho} - \dot{\gamma}'^{\nu]} X_{\rho} }{ ( -X^2)^{3/2} } \right) d\tau'
        \nonumber
        \\
        ~+ \frac{1}{2} ( \ddot{\gamma}^{\nu]} \dot{\gamma}_\rho - \dot{\gamma}^{\nu]} \ddot{\gamma}_\rho ) \ln (\epsilon/e \ell) \Bigg],
        \label{torque3D}
\end{align}
which depends both on a particle's charge $q$ and on its electromagnetic dipole moment $q^{\mu\nu}$.  It is clear in this context that varying $\ell$ changes the force only by constant multiples of $\ddot{\gamma}_\mu$. Different values for $\ell$ thus provide different renormalizations of a particle's apparent  mass, at least to leading nontrivial order.

Additional insight into our $d=3$ forces and torques may be gained by evaluating them in a slow-motion approximation. Applying such an approximation while integrating \eqref{eq:Lorentz3D} once by parts shows that the spatial 2-vector components of the self-force are explicitly
\begin{align}
        \hat{\bm{f}}(\tau) = - \frac{q^2}{2} \bigg[ \int_{-\infty}^{\tau-\epsilon}\! \! d\tau' \left( \frac{ \ddot{\bm{\gamma}}(\tau') }{ \tau-\tau' } \right) + \ddot{\bm{\gamma}}(\tau)  \ln \left( \frac{ \epsilon }{ e^{\frac{1}{2}} \ell} \right) \bigg],
        \label{Lorentz3Dslow1}
\end{align}
where it has been assumed that the acceleration falls off according to
\begin{equation}
        \lim_{\tau' \rightarrow -\infty} \frac{ (\tau - \tau') \dot{\bm{\gamma}}(\tau') - [ \bm{\gamma}(\tau) - \bm{\gamma}(\tau') ] }{ (\tau-\tau')^2 } = 0
        \label{fall1}
\end{equation}
in the distant past. If this falloff condition is indeed satisfied, the three-dimensional self-force thus depends on a weighted history of the charge's past acceleration. Beyond noting that the relevant weighting factor decays like $1/\tau$, the $\epsilon \rightarrow 0^+$ limit makes it difficult to interpret \eqref{Lorentz3Dslow1} directly. A manifestly-finite form for the self-force can be obtained by integrating by parts once more. Assuming that
\begin{equation}
        \lim_{\tau' \rightarrow -\infty} \ddot{\bm{\gamma}}(\tau') \ln (\tau-\tau') = 0,
        \label{fall2}
\end{equation}
the $d=3$ self-force may be seen to reduce to
\begin{equation}
        \hat{\bm{f}}(\tau) = -\frac{q^2}{2} \int_{-\infty}^\tau \!\! \dddot{\bm{\gamma}}(\tau')  \ln \left ( \frac{ \tau-\tau'}{ e^{\frac{1}{2}} \ell} \right)  d\tau' .
        \label{fNonrel}
\end{equation}
This depends on a past history of the particle's jerk, with a weighting factor which \textit{increases} logarithmically in the increasingly-distant past.

Similar expressions may be obtained for the slow-motion limit of the $d=3$ self-torque \eqref{torque3D}. If the falloff conditions \eqref{fall1} and \eqref{fall2} are assumed to hold and the electric and magnetic components of the particle's dipole moment may be considered comparable, the time-space components of the self-torque reduce to
\begin{equation}
        \hat{n}^{0i}(\tau) = q^{ij}(\tau) \hat{f}_j(\tau)/q,
\end{equation}
where $i,j \in \{ 1,2 \}$, the $\hat{f}_j$ appearing here is given by \eqref{fNonrel}, and we have assumed that $q \neq 0$. Differences between $\hat{\bm{p}}$ and $\hat{m} \dot{\bm{\gamma}}$ are thus controlled, in part, by the coupling of a particle's magnetic dipole moment to a logarithmically-weighted history of its jerk.

The remaining space-space components of the nonrelativistic $d=3$ self-torque, which directly affect a body's spin evolution, are determined by
\begin{equation}
        \hat{n}^{ij}(\tau) =  2 q^{0[i}(\tau) \hat{f}^{j]}(\tau)/q .
        \label{nij3D}
\end{equation}
The spin, which has only one component in this case, is thus affected by misalignments between a body's electric dipole moment and the same logarithmically-weighted history of its jerk.

Analogous expressions are more complicated when $d=5$. We give only the leading-order self-force, which is again found by substituting \eqref{W} and \eqref{DW} into \eqref{FhatFin}, and then substituting the effective field which results into \eqref{fLorentz}. The fully-relativistic force is thus
\begin{align}
        \hat{f}_\mu = - q^{2} \Bigg[ \int_{-\infty}^{\tau-\epsilon} \frac{  3 X_{[\mu} \dot{\gamma}_{\nu]}' \dot{\gamma}^\nu }{ (-X^2)^{3/2} } d\tau'+h_{\mu}{}^{\nu} \bigg( \frac{3\ddot{\gamma}_{\nu} }{8\epsilon^{2}} - \frac{\dddot{\gamma}_\nu}{2\epsilon} \nonumber\\
        ~ - \frac{3}{16} (\gamma^{(4)}_\nu - \frac{3}{2} |\ddot{\gamma}|^2 \ddot{\gamma}_\nu ) \ln ( \epsilon/e^{\frac{1}{3}} \ell) - \frac{|\ddot{\gamma}|^2 \ddot{\gamma}_\nu  }{32} \bigg) \Bigg].
        \label{eq:Lorentz5D}
\end{align}
Changing $\ell$ is this context may be seen to shift more than just the apparent mass; noting that
\begin{equation}
        h_{\mu}{}^{\nu} \left( \gamma_\nu^{(4)} - \frac{3}{2} |\ddot{\gamma}|^2 \ddot{\gamma}_\nu \right) = \frac{d}{d\tau} \left( \dddot{\gamma}_\mu - \frac{3}{2} |\ddot{\gamma}|^2 \dot{\gamma}_\mu \right),
\end{equation}
it affects both the direction and magnitude of the renormalized 5-momentum.

We note also that in a slow-motion limit, the spatial components of \eqref{eq:Lorentz5D} reduce to
\begin{equation}
        \hat{\bm{f}}(\tau) = \frac{3 q^2 }{16}  \int_{-\infty}^\tau \!\! \bm{\gamma}^{(5)}(\tau')  \ln \left ( \frac{ \tau-\tau'}{ e^{-\frac{17}{12}} \ell} \right)  d\tau' ,
        \label{fNonrel5D}
\end{equation}
at least if derivatives of the particle's position fall off sufficiently rapidly in the distant past. This differs from its $d=3$ counterpart \eqref{fNonrel} mainly by an overall sign and by the replacement of $\dddot{\bm{\gamma}}$ with $\bm{\gamma}^{(5)}$ in the integral.

\subsubsection{Comparisons}

We close this section by comparing with other odd-dimensional self-force results which have appeared in the literature. First, a five-dimensional force similar to \eqref{fNonrel5D} has recently been obtained using the methods of effective field theory \cite{Chad}. In that context, $\ell$ appears as a free parameter in a dimensional regularization procedure. This is not so different from our usage of $\ell$ as a free parameter in the choice of propagator used to define a body's momentum: Our propagator induces an $\ell$-dependent map $F_{ab} \mapsto \hat{F}_{ab}$, and this turns into an $\ell$-dependent regularization in the point particle limit. Nevertheless, we note that our results differ conceptually in that we have provided precise ``microscopic'' (or ``UV-complete'') definitions for the mass, mass center, other quantities appearing in the laws of motion; we do not merely assert that quantities satisfying such laws exist and that they have physical interpretations consistent with their names.

Other odd-dimensional self-forces which have appeared in the literature differ much more significantly from ours. These have been obtained by the use of heuristic arguments to directly regularize point-particle self-fields \cite{Galtsov, Kazinski}, expressions for the momenta associated with those fields \cite{Yaremko}, or similar quantities. In at least one case, the claimed force law is IR-divergent; see Eq. (4.4) in \cite{Galtsov}. Other proposals use counterterms which depend on a particle's entire past history \cite{Kazinski, Yaremko}, implying that a body's momentum could not be computed without knowledge of that history---a physically-unacceptable option. Another result predicts a time-varying mass even at leading order \cite{Yaremko}. While mass variations are normal and expected when including effects due to a body's higher multipole moments \cite{Dix70a}, they should not arise when considering only monopole interactions with an electromagnetic field. Indeed, it is clear from \eqref{fLorentz}, \eqref{pDot}, and \eqref{mDef} that mass variations do not arise in our leading-order expressions.

\section{Phenomenology of the odd-dimensional self-force}
\label{Sect:pheno}

Although the results of Section \ref{Sect:forceOdd} may be used to evaluate odd-dimensional point-particle self-forces and self-torques, the physical implications of those results are not immediately apparent. We now elucidate some of those implications, with a particular emphasis on non-relativistic systems in flat, three-dimensional spacetimes. This setting i) possesses features which are particularly distinct from the $d=4$ case, and ii) may find experimentally-accessible analogs in certain condensed-matter or fluid systems. Nevertheless no attempt is made here to provide a comprehensive discussion of $d=3$ self-force effects. Rather, we seek mainly to highlight some of the subtleties and unusual features of these effects.

\subsection{Approximations}
\label{Sect:approx}

We begin our discussion of odd-dimensional self-force phenomenology by remarking on some of the relevant approximations. Although we have already noted that the results of the previous section assume a type of point particle limit, the details of that limit were not fixed. Indeed, a number of different point particle limits can be consistently discussed, and without a specific physical system in mind, it is difficult to settle on a particular approximation. Despite this, one generic constraint which can be used is that physically-realisable bodies cannot exist with arbitrary combinations of physical size, charge, and mass. Energy conditions may be violated if the stresses required to counteract a body's internal electrostatic repulsion become larger than its mass density. Letting $L$ characterize a charge's linear dimension, those stresses might be estimated to be order $(q /L^{d-2})^2$. Noting that the mass density is approximately $m/L^{d-1}$, energy conditions thus demand that
\begin{equation}
        q^2 \lesssim m L^{d-3},
        \label{energyCond}
\end{equation}
where we have used the bare mass $m$ associated with the bare momentum $P_\tau$, which is defined by \eqref{PDef}. The renormalized mass $\hat{m}$ is however derived from $\hat{P}_\tau$, which is distinguished from $P_\tau$ via \eqref{Prenorm}. The bare and renormalized masses can  differ from one another by terms of order $q^2/L^{d-3}$ and $q^2 \ln (L/\ell)/L^{d-3}$. If $\ell$ is held fixed, saturating the bound in \eqref{energyCond} might then result in an ``imaginary $\hat{m}$,'' i.e., a spacelike $\hat{p}_a$. Other pathologies could arise as well. Our formalism breaks down in such cases, which we avoid by additionally requiring that
\begin{equation}
        q^2 \lesssim \frac{ m L^{d-3} }{ | \ln (L/\ell) | }.
        \label{qLog}
\end{equation}
This is sufficient to imply that $m$ and $\hat{m}$ have similar magnitudes.

If a charge moves in an externally-imposed electric field, the self and external forces acting on it may now be estimated to scale like
\begin{equation}
        f_\mathrm{self} \sim (q^2/\hat{m})\frac{ f_\mathrm{ext} }{ \tau_*^{d-3} } \lesssim \frac{ (L/\tau_*)^{d-3}  }{ |\ln(L/\ell)| } f_\mathrm{ext} ,
\end{equation}
where $\tau_*$ is a characteristic timescale associated with the external field. If $L$ is sufficiently small and $\tau_*$ is independent of $L$, self-forces thus remain at least logarithmically-suppressed in comparison with external forces, even for objects which are ``maximally charged'' according to \eqref{qLog}. We note however, that this statement is not precise. What meaning it does have is global, in the sense that nontrivial tails imply that self-forces can be instantaneously significant even when external forces vanish.

It would be interesting to now write down and systematically analyze the consequences of a complete, self-consistent approximation scheme which saturates the given bounds. We do not do so, however. Instead, we consider a simpler model problem in which only the mass and charge monopoles are significant. In this case, the momentum-velocity relation \eqref{momVel} reduces simply to $\hat{p}^a = \hat{m} \dot{\gamma}^a$ and the force is given entirely by the Lorentz term \eqref{fLorentz}. With these assumptions, \eqref{fNonrel} implies that the non-relativistic $d=3$ equation of motion is given by the integral equation
\begin{align}
        \hat{m} \ddot{\bm{\gamma}}(\tau) =&~ q \bm{E}_\mathrm{ext}( \bm{\gamma}(\tau) )
        \nonumber
        \\
        &~ -\frac{q^2}{2} \int_{-\infty}^\tau \!\! \dddot{\bm{\gamma}}(\tau')  \ln \left ( \frac{ \tau-\tau'}{ e^{\frac{1}{2}} \ell} \right)  d\tau' ,
        \label{EOM3D}
\end{align}
where $\bm{E}_\mathrm{ext}$ denotes the external electric field. Similarly, \eqref{fNonrel5D} implies that with the same assumptions, the $d=5$ equation of motion is
\begin{align}
        \hat{m} \ddot{\bm{\gamma}}(\tau) =&~ q \bm{E}_\mathrm{ext}( \bm{\gamma}(\tau) )
        \nonumber
        \\
        &~ + \frac{3 q^2}{16} \int_{-\infty}^\tau \!\! \bm{\gamma}^{(5)}(\tau')  \ln \left ( \frac{ \tau-\tau'}{ e^{-\frac{17}{12}} \ell} \right)  d\tau' .
        \label{EOM5D}
\end{align}
More systematic approximations would also include various test body effects involving the spin and higher-order electromagnetic multipole moments.

\subsection{Runaway solutions}
\label{Sect:Runaway}

The simplest applications for the equations of motion \eqref{EOM3D} and \eqref{EOM5D} concern the behavior of free particles. Unaccelerated trajectories are of course valid solutions when $\bm{E}_\mathrm{ext} = 0$, although they are not the only solutions. The space of possible initial data for these integral equations is infinite dimensional, and nontrivial choices for this data generically lead to nontrivial trajectories. Physically, this is as expected. However, there also exist solutions which are not physically reasonable. These ``runaway solutions'' accelerate exponentially and without bound: Letting $\bm{a}_0$ denote an arbitrary constant vector, suppose that
\begin{equation}
        \ddot{\bm{\gamma}}(\tau) = \bm{a}_0 \exp(\tau/\tau_\mathrm{run}).
        \label{aRun}
\end{equation}
If $d=3$, substitution of this expression into \eqref{EOM3D} shows that it is a solution when
\begin{equation}
        \tau_\mathrm{run} = \ell \exp(\gamma_\subE + 1/2 -2 \hat{m}/q^2 ),
        \label{tauRun}
\end{equation}
where $\gamma_\subE$ denotes the Euler-Mascheroni constant. Note that although $\tau_\mathrm{run}$ may appear to depend on the arbitrarily-chosen lengthscale $\ell$, the implicit dependence of $\hat{m}$ on $\ln \ell$ ensures that it does not.

More importantly, the existence of runaway solutions suggests that a particle upon which no force has been applied might spontaneously and violently accelerate without any apparent cause. One may hope that the runaway solutions are artifacts of the initial data (or lack thereof), in that solutions for which $\ddot{\bm{\gamma}}(\tau) = 0$ for all $\tau < \tau_0$ might behave more sensibly. Unfortunately, this is not so. The three-dimensional equation of motion \eqref{EOM3D} may be solved using Laplace transforms, and doing so shows that with trivial initial data, almost any applied force excites a runaway mode with growth timescale $\tau_\mathrm{run}$.

The situation is somewhat better when $d=5$. Substituting the ansatz \eqref{aRun} into \eqref{EOM5D}, the runaway timescale may be seen to satisfy
\begin{equation}
        \frac{ \hat{m} }{q^2}  = \frac{1}{64 \tau_\mathrm{run}^2} \left\{ 17 + 12 [ \ln (\tau_\mathrm{run}/\ell) - \gamma_\subE  ] \right\} .
        \label{runaway5D}
\end{equation}
However, the right-hand side here has a maximum when varying over all $\tau_\mathrm{run} > 0$, implying that runaway solutions can exist (with the given form) only when
\begin{equation}
        q^2 / \hat{m} \geq \frac{ 32 }{ 3 } \ell^2 \exp (2 \gamma_\subE-11/6) .
        \label{runawayBound}
\end{equation}
If this bound holds but is not saturated, there are in fact two solutions to \eqref{runaway5D}, and thus two runaway timescales. If the bound is violated, solutions to our equation of motion appear not to be unstable in five dimensions.

Although we are not aware of runaway solutions having previously been discussed in odd-dimensional spacetimes, they are well-known features of the $d=4$ Abraham-Lorentz-Dirac equation. One objection to them (besides their manifest disagreement with observation) is that their associated timescale is extremely short---of order $q^2/\hat{m}$ when $d=4$. However, \eqref{energyCond} implies that a well-defined four-dimensional point particle limit requires that $q^2/\hat{m}$ be of order $L$ or smaller. Additionally, standard derivations assume that all dynamical timescales are much longer than $L$. Runaway solutions in four dimensions are thus solutions to an equation whose properties violate the conditions under which that equation has been derived. In this sense, they are not genuine predictions.

Similar conclusions may be reached also when $d = 3$ or $d=5$; the runaway solutions discussed above cannot be considered genuine predictions of the theory. This is most easily seen in the five-dimensional case, for which \eqref{qLog} implies that $q^2 / \hat{m} \to 0^+$ in a point particle limit. The bound \eqref{runawayBound} is thus violated for sufficiently-small bodies, which means that runaway solutions do not exist in the relevant portion of parameter space. If $d=3$, runaway solutions do exist formally, although they violate the conditions under which the equation of motion may be expected to hold. A body which is maximally charged according to \eqref{qLog} has a runaway timescale \eqref{tauRun} which is short compared to its light-crossing time $L$, and particles with less charge have runaway timescales which are even shorter. However, our derivation breaks down for timescales of order $L$; runaway solutions are thus unphysical also in three dimensions.

\subsection{Reduction of order}
\label{Sect:reduced}

Although runaway solutions are not true predictions of our equations, it would be desirable to be able to systematically extract solutions which are physically and mathematically justified---well-behaved trajectories which are sufficiently close to satisfying, e.g., \eqref{EOM3D} and for which all significant timescales are much larger than $L$. By analogy with the $d=4$ case, we accomplish by ``reducing order,'' which corresponds to supposing that the external force alone generates a ``zeroth order'' trajectory determined by $\ddot{\bm{\gamma}} \approx q \bm{E}_\mathrm{ext}/\hat{m}$, and that it is this trajectory which should be substituted into the self-force integrals. If $d=3$, such a procedure results in
\begin{align}
                \hat{m} \ddot{\bm{\gamma}}(\tau)& =  q \bm{E}_\mathrm{ext}(\bm{\gamma}(\tau))
                \nonumber
                \\
                & ~-\frac{q^3}{2\hat{m}} \int_{-\infty}^\tau \! \! \dot{\bm{E}}_\mathrm{ext}(\bm{\gamma}(\tau'))  \ln \left ( \frac{ \tau-\tau'}{ e^{\frac{1}{2}} \ell} \right)  d\tau' .
                \label{EOMreduced3D}
\end{align}
When $d=5$, one finds instead that
\begin{align}
        \hat{m} \ddot{\bm{\gamma}}(\tau) &= q \bm{E}_\mathrm{ext}(\bm{\gamma}(\tau))
        \nonumber
        \\
        &~  + \frac{3q^3}{16\hat{m}} \int_{-\infty}^\tau \!\! \dddot{\bm{E}}_\mathrm{ext}(\bm{\gamma}(\tau'))  \ln \left ( \frac{ \tau-\tau'}{ e^{-\frac{17}{12}} \ell} \right)  d\tau'.
                \label{EOMreduced5D}
\end{align}
These replacements do not change the order of the approximation as long as $q^2/\hat{m}$ is sufficiently small. More to the point, they mollify the high-frequency character of the Fourier transforms associated with the unmodified accelerations (as is made more clear in Section \ref{Sect:ExactSoln} below). Regardless of justification, these equations no longer admit runaways and there is a sense in which their solutions nearly satisfy their parent equations as long as the self-force is sufficiently small. However, as we shall see below, the reduced-order equations can still be problematic when applied over very long times.

To briefly remark on our terminology, the reduction-of-order procedure applied to the $d=4$ Abraham-Lorentz-Dirac equation yields what is sometimes referred to as the Landau-Lifshitz equation. In that case, it has the mathematical effect of reducing the order of the relevant differential equation from three to two. Here, the reduced-order terminology is retained even though we are not changing the order of a differential equation.

We also note that the reduction-of-order procedure is not as \textit{ad hoc} as it might appear. It arises naturally when constructing more careful point particle limits; see \cite{GrallaHarteWald} at least for the $d=4$ case.

\subsection{Exact and approximate solutions without runaways}
\label{Sect:ExactSoln}

We next discuss how physically-acceptable exact and approximate solutions---i.e., solutions which do not run away---of the integro-differential equation of motion (\ref{EOM3D}) can be obtained when $d=3$, how the reduced-order equation \eqref{EOMreduced3D} arises in a certain limit, and how reduction of order breaks down over very long timescales.

First note that our original equation \eqref{EOM3D}, which assumes that the acceleration vanishes in the distant past, can be recast as
\begin{equation}
        \frac{2}{q} \bm{E}_\mathrm{ext} (\bm{\gamma}(\tau)) = \int_{-\infty}^\tau \!\! \dddot{\bm{\gamma}}(\tau') \ln \left( \frac{ \tau-\tau' }{\ell  \exp( \frac{1}{2} - \frac{ 2\hat{m} }{ q^2 } ) } \right) d\tau'.
\label{EOM3Da}
\end{equation}
This may be viewed as a linear integral equation for the particle's jerk $\dddot{\bm{\gamma}}$ in terms of the prescribed external force $q \bm{E}_\mathrm{ext}$. In particular, it is a Volterra equation of the first kind. Such equations are often solved using Laplace transforms. If the initial data is trivial, solutions obtained in this way generically display the runaway behavior mentioned above. However, there does exist nontrivial initial data for which no such problems arise. This data is selected automatically by using Fourier transforms instead of Laplace transforms, as the former cannot be used to represent exponentially-growing solutions. Indeed, we view the solution obtained by Fourier transform to be ``the'' physical one in a wide range of scenarios.

It is first convenient to define the body's acceleration as it would be in the absence of self-interaction:
\be
        {\bm a}_{\rm ext} \equiv \frac{q}{\hat{m}} {\bm E}_{\rm ext}.
        \label{aextdef}
\ee
Also defining the dimensionless time variable
\be
        s \equiv ( \tau / \tau_\mathrm{run}) e^{\gamma_\subE}
        \label{rescale1}
\ee
and its primed equivalent in terms of the runaway time \eqref{tauRun} and the Euler-Mascheroni constant $\gamma_\subE$, the body's true acceleration $\bm{a} = \ddot{\bm{\gamma}}$ is found from Eq.\ (\ref{EOM3Da}) to satisfy
\be
{\bm a}_{\rm ext}(s) = \frac{q^2}{2 \hat{m}} \int_{-\infty}^s \frac{ d {\bm a}}{ds'}(s')
\ln(s-s') ds'.
\label{ee1}
\ee
Since this equation is linear, a general solution can be written as
\be
{\bm a}(s) = \frac{2 \hat{m}}{q^2} \int_{-\infty}^\infty K(s-s')  {\bm a}_{\rm
  ext}(s') ds'
\label{ee2}
\ee
for some kernel $K$, where the factor $2 \hat{m}/q^2$ has been included for later convenience.

To solve for $K$, we now assume that the Fourier transform
of the solution exists. As mentioned above, this
assumption excludes runaway solutions, and so yields only a
certain class of solutions of the original equation. Defining the Fourier transform of the kernel by
\be
{\tilde K}(\omega) = \frac{1}{\sqrt{2 \pi}} \int ds \, e^{i \omega s}
K(s),
\ee
and substituting into (\ref{ee1}) and (\ref{ee2}), we find that
\be
{\tilde K}(\omega) = \frac{i}{2 \pi} \frac{1}{\omega {\tilde G}(\omega)},
\ee
where $G(s) \equiv \Theta(s) \ln (s)$. Evaluating the Fourier
transform of $G(s)$ now yields
\be
{\tilde K}(\omega) = - \frac{1}{\sqrt{2 \pi}} \frac{1}{ \ln_+(\omega
  e^{\gamma_\subE}) - i \pi/2},
\label{Kans}
\ee
where $\ln_+ (\omega)$ is the function obtained by analytically continuing
$\ln (\omega)$ from the positive real axis into the upper half
$\omega$ plane.  In particular, for real $\omega$, we have
\be
\ln_+(\omega) = \ln |\omega| + i \pi \Theta(-\omega).
\ee
One consequence is that
\begin{equation}
        \int_{-\infty}^\infty K(s) ds = \sqrt{2\pi} \tilde{K}(0) = 0.
        \label{zeroInt}
\end{equation}
In combination with \eqref{ee2}, it follows that with appropriate falloff conditions on $\bm{a}_\mathrm{ext}$,
\begin{equation}
        \Delta \bm{v} \equiv \int_{-\infty}^\infty \bm{a}(\tau) d\tau = 0.
\end{equation}
Initially-stationary particles thus return to rest at late times, an effect which is discussed further in Section \ref{Sect:Kicks} below.

A particle's motion at finite times can be understood by obtaining an expression for the kernel in the time domain, which of course follows from the inverse Fourier transform of (\ref{Kans}):
\be
K(s) = - \frac{1}{2 \pi} \int \frac{ e^{-i \omega s} d\omega }{
  \ln_+( \omega e^{\gamma_\subE} ) - i \pi/2 }.
  \label{KinvF}
\ee
We note that the Fourier transform ${\tilde
  G}(\omega) \propto 1/{\tilde K}(\omega)$ is analytic in the upper
half $\omega$ plane, which reflects the causal nature of $G(s)$:
\be
G(s) = 0, \qquad  s < 0.
\ee
By contrast, taking the reciprocal of $\tilde{G}(\omega)$ to find $\tilde{K}(\omega)$ results in a simple pole at
\be
\omega = i e^{-\gamma_\subE},
\label{pole}
\ee
indicating that the kernel $K(s)$ does not vanish for $s <
0$. The motion given by the solution (\ref{ee2}) thus exhibits a degree of
``preacceleration,'' just as for solutions of the Abraham-Lorentz-Dirac
equation in four dimensions. Preacceleration arises in both of these cases when one imposes that the solution does not
  diverge at late times. Although the three and four-dimensional equations of motion are mathematically quite different, such an imposition necessarily requires knowledge of the future---violating causality. We now show that this violation is confined to very small timescales which are effectively negligible.

For $s<0$, the inverse Fourier transform \eqref{KinvF} can be evaluated by completing the contour into a
semicircle in the upper half plane and evaluating the residue at the
pole (\ref{pole}), yielding
\be
K(s) = e^{-\gamma_\subE} \exp ( - e^{-\gamma_\subE} |s|
), \qquad s<0.
\label{acausal}
\ee
Although the kernel is acausal, its acausality is thus limited to a
specific timescale over which $s$ varies of order $e^{\gamma_\subE}$. Recalling \eqref{rescale1}, this corresponds to a physical timescale equal to the runaway time $\tau_\mathrm{run}$, given by \eqref{tauRun}. As argued in Section \ref{Sect:Runaway}, this timescale is short compared to the body's size $L$; it is negligible.

% However, one might still worry that the large factor $2 \hat{m}/q^2$ in \eqref{ee2} might amplify such effects. This is not the case: The acausal contribution to the acceleration predicted by that equation is
%\begin{equation}
%       \frac{ 2 \hat{m}/q^2 }{ \tau_\mathrm{run}} \int_\tau^\infty \bm{a}_\mathrm{ext}(\tau') \exp \left(- \left| \frac{ \tau - \tau'}{\tau_\mathrm{run} }\right| \right) d\tau'.
%\end{equation}

If $s>0$, one can instead complete the contour in \eqref{KinvF} into a semicircle in
the lower half $\omega$ plane, with a detour around branch cut at
${\rm Arg}(\omega) = -\pi/2$.  This yields an expression for the
kernel in the form of a Laplace transform
\be
K(s) = - \int_0^\infty \frac{  e^{-s \sigma} d\sigma }{\ln( \sigma
  e^{\gamma_E} )^2 + \pi^2 }, \qquad s > 0.
\label{Kpos}
\ee
While we have been unable to find an explicit analytic expression for
$K(s)$ for $s$ positive, it follows that an upper
bound is
\be
|K(s)| \le \frac{1}{\pi^2 s}
\ee
for all $s > 0$. This indicates that the memory of an external force on a body's acceleration decays at least as fast as $1/\tau$.

To summarize up to this point, we have found, for generic external fields, exact, physically-acceptable solutions to the $d=3$ equation of motion (\ref{EOM3D}). The accelerations corresponding to these solutions are given by (\ref{ee2}), where $\bm{a}_\mathrm{ext}$ is defined by (\ref{aextdef}), $s$ is defined by (\ref{rescale1}), and where $K(s)$ satisfies \eqref{acausal} and \eqref{Kpos}.

The Laplace transform expression (\ref{Kpos}) for the kernel $K(s)$ for $s>0$ is not very transparent.  We now develop a useful approximation to this kernel.  We have in mind two small quantities. First, the limiting process
discussed in Section \ref{Sect:approx} above requires that $q^2 \ll \hat{m}$. Second, we define $\tau_*$ to be a timescale over which the external electric field
varies, and define the dimensionless quantity $\nu$ by
\be
\nu^2 \equiv \tau_{\mathrm{run}} / \tau_*.
\label{nudef}
\ee
We assume $\nu$ to be small and throw away terms that are suppressed by one or more powers of it.

%the integrand of which has a local maximum at $u \sim 2/\ln s$
%and goes to zero at the two limits $u \to 0$ and $u \to \infty$.

An approximate expression for the kernel (\ref{Kpos}) at large $s$ can now be obtained as follows: Changing the variable of integration from $\sigma$ to $u = s \sigma$
we first obtain
\begin{equation}
K(s) = - \frac{1}{s} \int_0^\infty \frac{ e^{-u} du }{[\ln(u e^{\gamma_E}) -(\ln s)^2]^2 + \pi^2 }.
\label{Ktrans}
\end{equation}
Expanding the integrand here at large $\ln s$ gives
\begin{align}
K(s) = - \frac{1}{s (\ln s)^2} \int_0^\infty du  e^{-u}
\Bigg[ 1 + \frac{2 \ln (u e^{\gamma_\subE})}{\ln s}
\nonumber
\\
~ + \mathcal{O} \left( \left( \frac{\ln
      u}{\ln s} \right)^2 \right) \Bigg],
\label{FFA}
\end{align}
which is an approximation that breaks down both at large $u$ and at small $u$.  At
large $u$, the errors in the integrand become of order unity when $u
\agt s$, but because of the exponential suppression factor in the
integrand, the overall fractional corrections to the integral scale as $e^{-s}$, which we neglect. At small $u$, the errors in the integrand are
of order unity or larger for $u \alt 1/s$, and the corresponding
overall fractional corrections to the integral scale as the size of this region compared with the value $u=u_\mathrm{peak} \sim 2/\ln s$ at which the integrand in \eqref{Ktrans} takes its maximum value; they are of order
$$
\frac{ 1/s }{u_\mathrm{peak}} \sim \frac{ \ln s}{s}.
$$
Terms with this relative magnitude are also neglected here. Evaluating the integral (\ref{FFA}) thus gives
\be
K(s) = - \frac{1}{s (\ln s)^2} \left[ 1 + \mathcal{O} \left(\frac{1}{(\ln
      s)^2}\right)
 \right]
\label{Kapprox}
\ee
for $s > 0$.

We now argue that the asymptotic form (\ref{Kapprox}) of the kernel is
sufficient for deriving a useful explicit approximation for the acceleration (\ref{ee2}). We start by writing the latter expression in the form
\begin{align}
{\bm a}(s) = \frac{2 \hat{m}}{q^2} \left( \int_{-\infty}^{\bar{s}} + \int_{\bar{s}}^{\infty} \right) d s' \, K(s') \, {\bm a}_{\rm
  ext}(s - s') ,
\label{ee22}
\end{align}
for some parameter $\bar{s}$. Although this parameter is clearly arbitrary, we find it convenient to set
\be
        \bar{s} = 1/\nu = \sqrt{ \tau_* / \tau_\mathrm{run} } \gg 1.
\label{Deltas0}
\ee
This accomplishes two goals. First, it allows us to use
\be
        {\bm a}_{\rm ext}(s - s') = {\bm a}_{\rm ext}(s) [1 + \mathcal{O}(\nu)]
\label{aaa}
\ee
for $|s'| \lesssim \bar{s}$, at least if we are not too close to the boundary of the support of $\bm{a}_\mathrm{ext}$. Second, if \eqref{Kapprox} is used to approximate the kernel in the second integral in \eqref{ee22}, the relative error in doing so is bounded by $\varepsilon^2$, where
\be
        \varepsilon^{-1} \equiv \ln s_0 = \frac{\hat{m}}{q^2} + \ln \sqrt{ \frac{ \tau_* }{ \ell \exp(\frac{1}{2} + \gamma_\subE ) } } \approx \frac{\hat{m}}{q^2}.
\ee

The first integral in \eqref{ee22} can now be approximated by substituting \eqref{aaa} when $|s'| \lesssim \bar{s}$ and noting that contributions from larger negative values of $s'$ are exponentially suppressed due to \eqref{acausal}. Combining this with \eqref{zeroInt} and \eqref{Kapprox}, it follows that
\begin{align}
        {\bm a}(s) = \frac{2 \hat{m}}{q^2} \int_{\bar{s}}^\infty \!\! \frac{ d s' }{ s' (\ln s')^2 }
\left[ {\bm a}_{\rm ext}(s - s') - {\bm a}_{\rm ext}(s) \right]
\nonumber
\\
 ~ \times \left[ 1 +
  \mathcal{O}(\nu,\varepsilon^2) \right].
\label{ee24}
\end{align}
A somewhat simpler expression arises when integrating by parts, which yields
\begin{align}
{\bm a}(s) = \frac{2 \hat{m}}{q^2} \int_{\bar{s}}^\infty
\frac{d s' }{\ln s'}
\,
\frac{d {\bm a}_{\rm ext}
(s - s') }{ds}  \left[ 1 +
  \mathcal{O}(\nu,\varepsilon^2) \right]
\label{ee25}
\end{align}
if it is assumed that $\bm{a}_\mathrm{ext} (s) \to 0$ as $s \to \infty$. Note that the omission of the  $s' = \bar{s}$ boundary term in this expression, which is equal to
\begin{equation}
        -\frac{2 \hat{m}}{q^2} \left( \frac{\left[ {\bm a}_{\rm ext}(s - \bar{s}) - {\bm a}_{\rm ext}(s)
\right]}{\ln \bar{s} } \right)
\end{equation}
up to terms of relative order $\nu$ or $\varepsilon^2$, results in errors of order
\be
\frac{ d {\bm a}_{\rm ext}}{ds}
\frac{ \varepsilon^{-1} \bar{s}}{ \ln \bar{s} } \sim {\bm a}_{\rm ext}
\left( \frac{  \tau_\mathrm{run} }{ \tau_* } \right) \bar{s} \sim \bm{a}_\mathrm{ext} \nu.
\ee
This is absorbed into the overall $\mathcal{O}(\nu,\varepsilon^2)$ relative error in \eqref{ee25}. Using \eqref{rescale1}, our approximation \eqref{ee25} can finally be rewritten in terms of the physical time $\tau$: Letting $\bar{\tau} \equiv e^{-\gamma_\subE} \bar{s} \tau_\mathrm{run} = e^{-\gamma_\subE} \sqrt{ \tau_\mathrm{run} \tau_*}$ [which is not to be confused with the $\bar{\tau}$ defined by \eqref{tBar}],
\begin{align}
{\bm a}(\tau) =  \frac{q}{\hat{m}} \int_{ \bar{\tau} }^\infty \! d \tau'  \left(
\frac{ \dot{\bm{E}}_{\rm ext}  (\bm{\gamma}(\tau - \tau'))
}{ 1 + (q^2/2 {\hat m}) \ln ( \tau' /e^{\frac{1}{2}} \ell
  ) } \right)
  \nonumber
  \\
~ \times \left[ 1 +
  \mathcal{O}(\nu,\varepsilon^2) \right].
\label{ee27}
\end{align}
This is our approximate solution to the $d=3$ equation of motion \eqref{EOM3D}.

The reduced-order equation \eqref{EOMreduced3D} can now be obtained directly from \eqref{ee27} by assuming that $\bm{E}_\mathrm{ext}(\bm{\gamma}(\tau))$ is nonzero only for a finite time, which we assume to be short compared to the timescale
\be
 \ell \exp(2 {\hat m}/q^2) \gg \ell \gg L.
\label{tlarge}
\ee
If, furthermore, we evaluate $\bm{a}(\tau)$ at times $\tau$ which are small
compared to this timescale, we can expand the denominator in
(\ref{ee27}) in a Taylor series in $q^2/2 \hat{m}$.
This yields
\begin{align}
{\bm a}(\tau) = \Bigg\{ {\bm a}_{\rm ext}(\tau)
- \frac{q^2}{2 {\hat m}}\int_{-\infty}^\tau
d \tau' \dot{\bm{a}}_{\rm ext} (\tau')  \ln \left( \frac{ \tau-\tau' }{e^{ \frac{1}{2} } \ell}
  \right)
\nonumber
\\
~ \times \left[ 1 +
  \mathcal{O} \left(\frac{q^2}{2 {\hat m}} \ln \left( \frac{ \tau}{\ell}
    \right) \right) \right] \Bigg\} \left[ 1 + \mathcal{O}(\nu, \varepsilon^2) \right],
\label{ee28}
\end{align}
where we have used \eqref{aaa} and also the fact that the lower limit of $\bar{\tau}$ in \eqref{ee27} can be replaced by a lower limit of $0$ while incurring relative errors only of order $\nu$. This result coincides with the expression (\ref{EOMreduced3D}) obtained earlier by reduction of order. At times large compared to the timescale (\ref{tlarge}), the approximation (\ref{ee28}) is no longer
valid, and one must instead use the original expression (\ref{ee27}).

\subsection{Special types of motion}

Our equations of motion may now be used to answer at least two types
of questions:
\begin{enumerate}
\item How does a small charge move in response to a
given external field?
\item Which external field is required in order
for a charge to move on a given trajectory?
\end{enumerate}
The first of these questions cannot generally be answered using
the exact equations of motion (\ref{EOM3D}) and (\ref{EOM5D}), since
their solutions generically involve
unphysical instabilities as discussed in Section \ref{Sect:Runaway}
above.
However, at least if $d=3$, one can instead use the reduced-order
equation (\ref{EOMreduced3D}) over short timescales, or more generally
(\ref{ee27}) over all timescales. Either of these possibilities yield approximate
solutions with no unphysical instabilities.

The second potential question we can address, concerning the force required to hold a particle on a
given trajectory, can be computed using either \eqref{EOM3D} or
\eqref{EOMreduced3D} when $d=3$, although it is the former unmodified equation
which is typically simpler for this purpose. Answers will in any case
be similar using either method, at least if all timescales associated
with the given trajectory are sufficiently long and $q^2/\hat{m}$ is
sufficiently small. We now discuss some simple examples.

\subsubsection{Exponential growth}

Our first case is that of exponential motion: Consider trajectories with the form \eqref{aRun}, where $\tau_\mathrm{run}$ is now replaced by a generic positive constant $\tau_*$. At least formally, \eqref{EOM3D} predicts that if $\tau_* = \tau_\mathrm{run}$, no external force is required to effect such a trajectory when $d=3$. A rather different prediction follows, however, from the reduced-order equation \eqref{EOMreduced3D}. If $\tau_* \gg \tau_\mathrm{run}$, both equations predict similar results; the unmodified one gives the exactly-exponential external force
\begin{equation}
        q \bm{E}_\mathrm{ext} = \left[ 1 + \frac{ q^2 }{ 2 \hat{m} } \left( \ln (\tau_*/\ell) - \frac{1}{2} - \gamma_\subE \right) \right] \hat{m} \bm{a}_0 e^{\tau/\tau_*}
\end{equation}
in three dimensions, while the reduced-order equation implies that if this force is applied, the particle's acceleration will be
\begin{equation}
        \ddot{\bm{\gamma}} = \left\{ 1 - \left[ \frac{ q^2 }{ 2 \hat{m} } \left( \ln (\tau_*/\ell) - \frac{1}{2} - \gamma_\subE \right)\right]^2 \right\} \bm{a}_0 e^{\tau/\tau_*}.
\end{equation}
The relative difference between this and our starting ansatz \eqref{aRun} is of order $(q^2/\hat{m})^2$, as expected when comparing equations in which order reduction has and has not been applied.

\subsubsection{Harmonic motion}

A more interesting example which can be understood analytically (and is mathematically similar) is that of harmonic motion. Suppose that the trajectory is given by
\begin{equation}
        \bm{\gamma}(\tau) = \Re [ \bm{\gamma}_0 \exp( i \omega \tau)],
\end{equation}
where $\omega$ is real and the constant vector $\bm{\gamma}_0$ may be complex. Such an acceleration violates the falloff condition \eqref{fall2} but not the weaker condition \eqref{fall1}. We therefore substitute into \eqref{Lorentz3Dslow1} to find that the leading-order external force required to maintain harmonic motion is
\begin{align}
        q \bm{E}_\mathrm{ext} = \left[ 1 - \frac{q^2}{2\hat{m}} \left( \ln |\omega|\ell + \frac{1}{2} + \gamma_\subE \right) \right] \hat{m}\ddot{\bm{\gamma}}
        \nonumber
        \\
        ~ + \frac{\pi}{4} q^2 |\omega| \dot{\bm{\gamma}}
        \label{fHarmonic}
\end{align}
when $d=3$. The analogous $d=5$ expression is very similar except for an additional overall factor of $\omega^2$ in the self-interaction terms. Regardless, if the motion is confined to one spatial dimension, the self-force acts to provide i) a damping force, and ii) a $\ln |\omega| \ell$ shift to a charge's apparent inertia. If the motion is instead circular, similar interpretations apply, except that it is only the component of the self-force which is proportional to the velocity that performs work.

\subsubsection{Power laws and analytic trajectories}

Another example which is easily understood is one in which the acceleration vanishes for all $\tau < \tau_0$, while
\begin{equation}
        \ddot{\bm{\gamma}}(\tau) = \bm{a}_n [(\tau-\tau_0)/\tau_*]^n
        \label{aPower}
\end{equation}
thereafter, where $\bm{a}_n$, $\tau_0$, $\tau_*$, and $n$ are constants (the latter two of which are assumed to be positive). Substituting this into \eqref{EOM3D} shows that the external force required to produce such an acceleration has a somewhat-different time dependence than the acceleration itself: In terms of the harmonic number $H_n$,
\begin{equation}
        q \bm{E}_\mathrm{ext} = \left\{ 1 + \frac{ q^2 }{ 2 \hat{m}} \left[ \ln \left( \frac{ \tau - \tau_0 }{ e^{\frac{1}{2}} \ell } \right) - H_n \right] \right\} \hat{m} \ddot{\bm{\gamma}}
        \label{fPower}
\end{equation}
when $d=3$ and $\tau > \tau_0$. The logarithm here implies that even at late times, there remains a strong ``memory''  of the ``turn-on event'' at $\tau = \tau_0$.

This result allows us to understand which external forces are needed to hold a charge on a more-general trajectory which is analytic for all all $\tau > \tau_0$. Suppose that  $\ddot{\bm{\gamma}}(\tau) = 0$ for $\tau< \tau_0$ and
\begin{equation}
        \ddot{\gamma}(\tau) = \sum_{n=1}^\infty \bm{a}_n [(\tau-\tau_0)/\tau_*]^n
\end{equation}
when $\tau \geq \tau_0$, where the $\bm{a}_n$ are constants. Combining \eqref{aPower} and \eqref{fPower}, the required external force is seen to be
\begin{align}
        q \bm{E}_\mathrm{ext}(\bm{\gamma}(\tau)) = \left[ 1 + \frac{ q^2 }{ 2 \hat{m}} \ln \left( \frac{ \tau - \tau_0 }{ e^{\frac{1}{2}} \ell } \right) \right]\hat{m} \ddot{\bm{\gamma}}(\tau)
        \nonumber
        \\
        ~ - \frac{q^2}{2} \sum_{n=1}^\infty \bm{a}_n H_n [(\tau-\tau_0)/\tau_*]^n.
\end{align}

\subsubsection{Kicks}
\label{Sect:Kicks}

Our last---and most interesting---example is concerned with a charge which is briefly ``kicked'' by some external force. Focusing again on three dimensions, we initially ask which external field must be imposed in order for a particle to be only momentarily accelerated: Suppose that a charge is initially stationary, is subjected to a brief acceleration near $\tau=\tau_0$, and moves inertially thereafter with velocity $\dot{\bm{\gamma}}(\tau) = \Delta \bm{v}$. Substituting this into \eqref{fNonrel} shows that the self-force at late times must be balanced by an external force satisfying
\begin{equation}
        q \bm{E}_\mathrm{ext}(\bm{\gamma}(\tau)) = \frac{q^2}{2} \left( \frac{\Delta \bm{v} }{ \tau-\tau_0 } \right),
        \label{fKick}
\end{equation}
where we have neglected terms of order $1/(\tau-\tau_0)^2$. The self-force thus acts to push the particle back towards rest. This effect persists indefinitely, suggesting that the particle's initially-stationary state creates a ``preferred rest frame'' to which it always attempts to return.

We now change perspective, asking not for the external force required to maintain a briefly-accelerated trajectory, but instead for the trajectory of a particle in which the external force is only briefly nonzero. There are potential physical issues associated with this scenario, essentially because it is not clear if the strong tails present in three dimensions preclude any possibility of setting up a prescribed, confined electric field; there may be unavoidable and significant remnants of the process by which any experiment might be assembled. See, e.g., \cite{WaldMemory} for some recent remarks---in a somewhat different context---on persistent memory effects in odd dimensions. Regardless, there is no mathematical difficulty with assuming a prescribed external field and we proceed without further comment.

\begin{figure}
        \includegraphics[width=.97\linewidth]{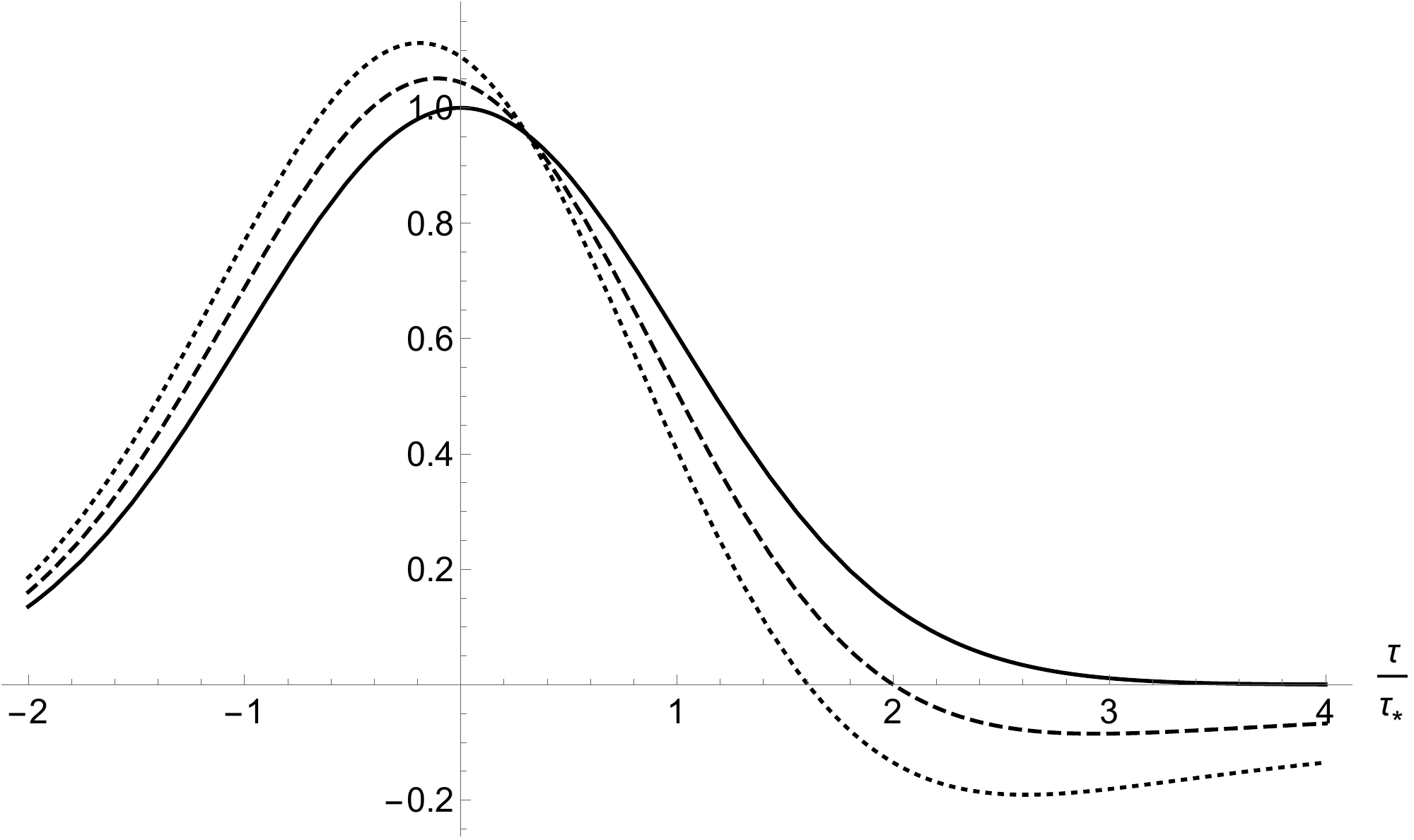}
        \caption{Net force as a function of $\tau/\tau_*$ for a Gaussian external field proportional to $\exp(-\frac{1}{2} (\tau/\tau_*)^2)$, as computed using the reduced-order $d=3$ equation of motion \eqref{EOMreduced3D}. Here, $\ell = \tau_* > 0$ and all results are normalized so that the maximum external force is equal to unity. The solid line corresponds to the external force, the dashed line to the net force when $q^2/2\hat{m} = 1/10$, and the dotted line to the net force when this parameter is equal to $1/5$. }
        \label{Fig}
\end{figure}

The net force acting on a charge for which the external field has a Gaussian profile is plotted in Figure \ref{Fig}, assuming the reduced-order equation of motion \eqref{EOMreduced3D}. Self-interaction is seen to slightly increase the peak magnitude of the force in this case, and also to shift that peak earlier in time. That the peak of the net force appears to anticipate the peak of the applied force might initially appear to violate causality, and to be reminiscent of the preacceleration seen in the Abraham-Lorentz-Dirac equation (and in the $d=3$ results discussed in Section \ref{Sect:ExactSoln} above). Causality is not violated here, however. The result arises from the explicitly-causal integral in \eqref{EOMreduced3D}, and appears because the self-force is sensitive to $\dot{\bm{E}}_\mathrm{ext}$, which decreases near the peak of the external force.

One can also see in the figure that the self-force eventually switches sign and only slowly returns to zero. A charge thus continues to decelerate long after the external field decays away. The late-time behavior of this process does not depend on whether or not the external field is Gaussian, and we now analyze more generally the asymptotic motion of a kicked charge.

Long after a briefly-nonzero external force has been applied, the reduced-order equation \eqref{EOMreduced3D} would suggest that the acceleration decays like $1/(\tau-\tau_0)$. However, an acceleration which decays this slowly implies a velocity which \textit{grows} logarithmically at late times. Such growth is unphysical. It may be traced back to a failure of the order reduction procedure at late times; cf. the derivation of \eqref{ee28} from \eqref{ee27}.

A more careful analysis using the methods of Section \ref{Sect:ExactSoln} shows that in fact, a particle asymptotically returns to its initial ``pre-kick'' velocity; see \eqref{zeroInt}. In essence, this recovers the Aristotelian idea that perturbed masses
eventually return to rest when all perturbations are removed. More precisely, (\ref{ee27}) shows that the asymptotic velocity of a
particle which is initially at rest decays like
\be
\dot{\bm{\gamma}}(\tau) = \frac{ \Delta \bm{v}}{1 + (q^2/2 {\hat
      m}) \ln [ (\tau-\tau_0)/(e^{\frac{1}{2}} \ell) ]}
\ee
at late times, where $\Delta \bm{v}$ is the time integral of ${\bm a}_{\rm
  ext}(\tau) = (q/\hat{m}) \bm{E}_\mathrm{ext} (\gamma(\tau))$. 

\section{Discussion}
\label{Sect:Discuss}

We have developed a general formalism with which to understand the motion of extended, self-interacting charges in all spacetime dimensions $d \geq 3$. Before understanding how such objects move, it is first necessary to fix precisely what should be meant by the concept of motion. We do so by giving precise definitions for a body's linear and angular momenta. One of the central properties of the momenta introduced here is that their laws of motion are structurally identical to the laws of motion satisfied by extended test bodies. This statement holds to all multipole orders, and for both an object's translational and rotational degrees of freedom. For example, the lowest-order force is given by the usual Lorentz expression \eqref{fLorentz}, and the lowest-order torque by \eqref{nDipole}. The only difference between these results and their test body counterparts is that the field $\hat{F}_{ab}$ which appears in them is a certain nonlocal linear transformation of the physical electromagnetic field $F_{ab}$. It is in the details of this field that the most visible effects of self-interaction may be found. Note as well that it is the same effective field which appears in expressions for both forces and torques, and that the prescription for this field remains the same at all multipole orders.

To be somewhat more precise, we do not find only a single momentum definition which obeys laws of motion structurally identical to test-body laws, but rather a class of such definitions. Different elements of this class become distinct only when self-interaction is significant, and they may be characterized by a certain 2-point ``propagator'' $G_{aa'}(x,x')$; see \eqref{PDef}, \eqref{pSDef}, and \eqref{Prenorm}. Physically, this propagator fixes a sense in which a charge element at $x'$ can source a field at $x$ whose net effect on the body's motion may be removed by finite renormalization of its multipole moments. We show from first principles that any propagator which satisfies the four ``axioms'' given in Section \ref{Sect:genDW} has this interpretation, and that such propagators may be used to define momenta with physically-desirable properties. Our axioms generalize the three originally proposed by Poisson \cite{PoissonLR} (in a somewhat different context) in order to characterize the $d=4$ propagators originally constructed by Detweiler and Whiting \cite{DetweilerWhiting2003}.

The axioms we introduce are essential to understanding the odd-dimensional self-force, and can be useful also in certain even-dimensional scenarios. However, they do not single out a unique propagator. Consequently, we do not have a unique momentum, a unique effective field, or even unique multipole moments associated with a body's stress-energy tensor. All of these quantities may depend on the choice of propagator. Nevertheless, such differences do not signal any kind of physical inconsistency. They merely reflect that one can choose to focus on slightly different aspects of the same physical system, and there is no reason to expect that all such aspects behave identically. A somewhat simpler ``gauge freedom'' of this kind arises even in $d=4$ discussions of extended test bodies, wherein different spin supplementary conditions may be applied to yield distinct centroids which nevertheless describe different aspects of the same physical system \cite{CostaReview}.

Having established an appropriate class of propagators with which to construct physically-useful momenta, it is essential to be able to find explicit examples in that class. In even numbers of dimensions, a straightforward generalization of the Detweiler-Whiting ``$S$-type'' Green function satisfies our constraints, and may therefore be used to generate suitable momenta for extended charge distributions. Adopting such definitions, the laws of motion involve effective electromagnetic fields which locally satisfy the source-free Maxwell equations. Extended self-interacting charges in even numbers of dimensions may thus be viewed as obeying laws of motion which are structurally identical to those of extended test bodies, and where the effective field appearing in those laws is source-free. This is a relatively straightforward generalization of existing $d=4$ results on relativistic motion in generic spacetimes \cite{HarteEM}. It may also be viewed as a generalization of the well-known statement that massive bodies interacting via Newtonian gravity or electrostatics satisfy laws of motion which involve only source-free external fields.

The odd-dimensional case is different. One of our main results is the identification of an odd-dimensional propagator, namely \eqref{EqLogProp0}, which satisfies the four constraints given in Section \ref{Sect:genDW}. This propagator is quite different from its even-dimensional Detweiler-Whiting counterpart; it is not a Green function or even a more general parametrix for Maxwell's equations. The effective field which appears in the laws of motion may thus fail to satisfy the source-free Maxwell equations when $d$ is odd. This difference is reasonably subtle at lower multipole orders. However, it may be qualitatively important when higher-order extended-body effects become significant: In that context, all components of a body's multipole moments may affect its motion, rather than only their (more familiar) trace-free components.

Another interesting feature of the odd-dimensional effective fields identified here is that in a point-particle limit, the map which translates the physical field into the effective field appears to turn into a kind of dimensional regularization procedure. This procedure arises as the limit of a map which is generically non-singular, makes no symmetry assumptions, and applies in a single spacetime with fixed integer dimension. A better understanding of this link may provide an improved understanding of dimensional regularization more generally.

Regardless, whether in even numbers of dimensions or odd, our formalism can be applied together with point particle limits to generate explicit laws of motion. We do so in Section \ref{Sect:Flat}, restricting to flat spacetimes for simplicity. Assuming retarded boundary conditions, we provide the general prescription for all dimensions $d \geq 3$, and apply it in full to find leading-order point-particle self-forces for $d=3,4,5,6,8$, and  leading-order self-torques for $d=3,4,5$. Our explicit self-forces agree with existing results in the literature when $d = 4,6,8$, although for $d \neq 4$, our approach is more systematic and includes microscopic definitions which were previously lacking. The odd-dimensional cases are different, and we identify significant problems with most other proposals which have been suggested in that context.

Finally, Section \ref{Sect:pheno} analyzes solutions to the nonrelativistic limits of our $d=3$ and $d=5$ results. The particularly-slow decay of odd-$d$ fields---particularly in three dimensions---results in a very strong dependence on a charge's past history: It follows from \eqref{fNonrel} that the self-force acting on a particle in $2+1$ dimensional Minkowski spacetime depends on the past history of its jerk $\dddot{\bm{\gamma}}(\tau)$,  with a weighting factor which \textit{grows} logarithmically in the increasingly-distant past.

Some physical consequences of this can be illustrated by considering a
charge which is briefly kicked by an externally-imposed electric field in a $d=3$ Minkowski spacetime. If this external field is Gaussian, one sees from Figure \ref{Fig} that self-interaction causes the peak of the net force
to arrive before the peak of the applied force. Despite appearances, this effect is causal. Moreover, for any external force---whether Gaussian or not---we show that if a charge is stationary for all time before an 
external field is applied, the slowly-decaying remnant of its self-field causes that charge to asymptotically return to rest at
late times. The slow decay of the self-field a body produces while it is initially at rest in three spacetime dimensions thus provides a preferred, dynamically-produced rest frame which persists and remains significant even in the distant future.

We note that although this paper has focused on the motion of bodies coupled to electromagnetic fields, our analysis extends straightforwardly for other types of interactions. For example, our odd-dimensional electromagnetic propagator \eqref{EqLogProp0} is replaced by
\begin{equation}
        G_\mathrm{odd} = \frac{ (-1)^{\frac{1}{2}-\lambda_d} U }{ 2\pi }  \lim_{\lambda \rightarrow \lambda_d}  \ell^{2\lambda}\frac{\partial}{\partial \lambda} \left[ ( 2 \sigma / \ell^2 )^\lambda \Theta (\sigma)\right],
\end{equation}
for a body coupled to a Klein-Gordon field in an odd-dimensional spacetime, where $U$ is a smooth biscalar which also appears in the retarded Green function. Furthermore, point-particle scalar fields can be obtained directly from our electromagnetic vector potentials by replacing the $W_a(x,\tau;\lambda)$ given by \eqref{WDef} with
\begin{equation}
        W(x,\tau; \lambda) = \frac{ q(\tau) }{ \alpha_d }  U(x,\gamma(\tau)) \Sigma^\lambda(x,\tau).
\end{equation}
We note as well that our methods generalize almost as easily for bodies coupled to (at least the linearized) $d$-dimensional Einstein equation.

As a simple application in the scalar setting, we note that masses can vary here even at monopole order, and that charges which source Klein-Gordon fields are not necessarily conserved. If an initially-uncharged body rapidly acquires a net charge $q_\infty$ around $\tau=\tau_0$, our equations show that the mass ``evaporates'' according to
\begin{equation}
        \hat{m}(\tau) - \hat{m}(\tau') = q_\infty^2 \ln \left(  \frac{ \tau' - \tau_0 }{ \tau - \tau_0 } \right)
\end{equation}
for a stationary particle in a $d=3$ Minkowski spacetime, where $\tau, \tau' \gg \tau_0$. Accounting for differences in numerical conventions, this matches an earlier result \cite{Burko2002} obtained using different methods. It is also conceptually similar to the scalar charge evaporation found for freely-falling charges in $d=4$ de Sitter spacetimes \cite{HartePoisson}.

Whether in electromagnetic or other contexts, there are various directions in which the results presented in this paper may be extended or applied. One possibility would be to relax our assumptions regarding retarded boundary conditions and trivial topology. Some discussion of motion in  topologically-nontrivial spacetimes has already been given \cite{StarkmanTopology}, although mainly in cases where the formally-divergent portion of the point-particle self-field could be clearly seen not to contribute to the self-force. The formalism developed here lays the groundwork for extending these kinds of results for generic types of motion: All of the formalism developed in Section \ref{Sect:NonPert} holds regardless of boundary or initial conditions, or topology, and the $S$-fields given by \eqref{eq:SingFieldEven} and \eqref{ASfin} are similarly-agnostic to these features. If a physical field $F_{ab}$ can be computed in some physical system---whether by numerical, perturbative, or other methods---the $S$-fields given here can be used to straightforwardly determine the force. One motivation for such generalizations is the potential for connecting this work with the behavior of certain lower-dimensional condensed matter systems, systems which are often characterized by nontrivial boundary conditions or topology. Moreover, experimental work in pilot-wave hydrodynamics \cite{PilotWave} suggests---although the mathematics applicable there is not precisely analogous to ours---that self-interaction problems in two spatial dimensions can have very rich and surprising behavior in the presence of nontrivial boundary conditions.

\appendix

\section{Propagators and Hadamard series}
\label{app:Hadamard}

This appendix explains how to determine the bitensors which appear in the even-dimensional Detweiler-Whiting Green functions $G_{aa'}^\subDW$ with the form \eqref{evenD}, and also in the odd-dimensional propagators $G_{aa'}^\mathrm{odd}$ given by \eqref{EqLogProp0}. In both cases, it is convenient to introduce the van Vleck determinant $\Delta(x,x')$, which is a symmetric biscalar satisfying \cite{PoissonLR, HFTPaper1}
\begin{equation}
        \sigma^a \nabla_a \ln \Delta = d- \nabla^a \nabla_a \sigma,
        \label{DeltaDef}
\end{equation}
and also $\Delta(x,x) = 1$. In general, $\sigma^a \equiv \nabla^a \sigma(x,x')$ lies tangent to the geodesic which passes through $x$ and $x'$, so \eqref{DeltaDef} may be viewed as a first-order ordinary differential equation for $\Delta(x,x')$ along that geodesic. If Synge's function is known, $d- \nabla^a \nabla_a \sigma$ is easily computed and the solution can be written as an explicit integral along that geodesic. Integral solutions for this and similar ``transport equations'' may be found in, e.g., Appendix B of \cite{HFTPaper1}.

\subsection{Even-dimensional propagators}
\label{app:Hadamardeven}

The even-dimensional Detweiler-Whiting Green function $G^\subDW_{aa'}$ involves two bitensors, $U_{aa'}$ and $V_{aa'}$. Substituting its form \eqref{evenD} into \eqref{Ggreen} shows that these must satisfy
\begin{align}
        0  = [ 2 \sigma^b \nabla_b U_{aa'} + ( \nabla^b \nabla_b \sigma - d ) U_{aa'} ] \delta^{(d/2-1)} (\sigma)
        \nonumber
        \\
        ~ + [\nabla^b \nabla_b U_{aa'} - R_{a}{}^{b} U_{ba'} ] \delta^{(d/2-2)}(\sigma)
        \nonumber
        \\
        ~ + [ 2 \sigma^b \nabla_b V_{aa'} + ( \nabla^b \nabla_b \sigma - 2 ) V_{aa'} ] \delta(\sigma)
        \nonumber
        \\
        ~ + [\nabla^b \nabla_b V_{aa'} - R_{a}{}^{b} V_{ba'}] \Theta(\sigma)
        \label{GDWBitensors}
\end{align}
when $x \neq x'$, and also
\begin{equation}
        \lim_{x' \to x} U_{ab'}(x,x') = \alpha_d g_{ab},
\end{equation}
where $\alpha_d$ is given by \eqref{alphaEven}. The first three lines restrict $U_{aa'}$ and $V_{aa'}$ on the $\sigma = 0$ light cones, while the last requires that $V_{aa'}$ satisfy the homogeneous Maxwell equation
\begin{equation}
        \nabla^b \nabla_b V_{aa'} - R_{a}{}^{b} V_{ba'} = 0,
        \label{BoxV}
\end{equation}
at least when $\sigma > 0$. We note that the bitensors determined by these equations also arise in the retarded and advanced Green functions, via
\begin{equation}
        G_{aa'}^{\ret,\mathrm{adv}} = [U_{aa'} \delta^{(d/2-2)} (\sigma) - V_{aa'} \Theta (-\sigma)]_{\ret,\mathrm{adv}},
        \label{GretEven}
\end{equation}
although they may be evaluated at different locations here than in $G^\subDW_{aa'}$.

In order to complete the solution to \eqref{GDWBitensors}, it is first convenient to factor out the square root of the van Vleck determinant and to expand in the Hadamard series
\begin{equation}
        U_{aa'} = \Delta^{1/2} \sum_{n=0}^{d/2-2} \frac{\sigma^n}{n!} \mathsf{U}_{aa'}^{\{n\}}  .
        \label{Uexpand}
\end{equation}
Note that this is not a Taylor expansion; the ``coefficients'' $\mathsf{U}_{aa'}^{\{n\}}$ may be nontrivial functions of $x$ and $x'$. Regardless, using \eqref{DeltaDef} and the identity
\begin{equation}
        \sigma^n \delta^{(p)}(\sigma) = \frac{ (-1)^n p! }{ (p-n)! } \delta^{(p-n)}(\sigma),
\end{equation}
while setting to zero explicitly-equal numbers of derivatives of $\delta(\sigma)$, we find that
\begin{equation}
        \mathsf{U}^{ \{0\} }_{aa'} = \alpha_d g_{aa'},
        \label{U0}
\end{equation}
and that for all $n \in \{ 1, \ldots, d/2-2\}$,
\begin{align}
        (\sigma^b &\nabla_b + n) \mathsf{U}^{ \{ n \} }_{aa'} = \frac{ n }{ d - 2 - 2n}
        \nonumber
        \\
        & ~ \times [ \Delta^{-1/2} \nabla^b \nabla_b (\Delta^{1/2} \mathsf{U}_{aa'}^{ \{ n-1 \} } ) - R_{a}{}^{b} \mathsf{U}_{ba'}^{ \{ n-1 \} } ].
        \label{Un}
\end{align}
These constitute a tower of transport equations for each $\mathsf{U}_{aa'}^{ \{n\} }$ in terms of $\mathsf{U}_{aa'}^{ \{n - 1\} }$. The lone nonsingular solutions to these differential equations are the physical ones. They guarantee that the first two lines of \eqref{GDWBitensors} vanish.

The last line of that equation vanishes by \eqref{BoxV}, while the third can be eliminated by imposing the ``boundary condition''
\begin{align}
        [\sigma^b & \nabla_b  +(d/2-1)] (\Delta^{-1/2} V_{aa'}) = \frac{(-1)^{d/2-1}}{2}
        \nonumber
        \\
        &~ \times [ \Delta^{-1/2} \nabla^b \nabla_b (\Delta^{1/2} \mathsf{U}_{aa'}^{ \{ d/2-2 \} } ) - R_{a}{}^{b} \mathsf{U}_{ba'}^{ \{ d/2-2 \} } ]
        \label{Vtrans}
\end{align}
on $V_{aa'}$ when its arguments are null-separated. Eqs. \eqref{BoxV}, \eqref{U0}, \eqref{Un}, and \eqref{Vtrans} together provide a complete solution to \eqref{GDWBitensors}, and thus a complete determination of $U_{aa'}$ and $V_{aa'}$.

Note that unlike when finding these bitensors for the retarded or advanced Green functions, solving \eqref{BoxV} with boundary data \eqref{Vtrans} constitutes a peculiar type of ``exterior'' characteristic problem: Data is specified on the past and future light cones and we seek a solution to the wave equation outside of those light cones. Although the general mathematical status of such problems is not particularly clear, a Hadamard-like series analogous to \eqref{Uexpand} can be developed for $V_{aa'}$, resulting in an infinite tower of transport equations for the Hadamard coefficients $\mathsf{V}_{aa'}^{\{n\}}$. We assume $V_{aa'}$ to be specified in this sense, and that the resulting series is well-behaved. In fact, it can be acceptable to use Detweiler-Whiting propagator in which the Hadamard series for $V_{aa'}$ is truncated at some finite order. Forces and torques due to the associated effective field would be slightly altered by this truncation, although that would be due to them describing rates of change of slightly different quantities; such propagators still  generate correct and useful laws of motion.

It is clear from this discussion that since each Hadamard coefficient $\mathsf{U}_{aa'}^{\{n\}}$ or $\mathsf{V}_{aa'}^{\{n\}}$ can be written as a line integral along the geodesic segment which connects its arguments, it can depend on the geometry only on that geodesic. This establishes that each coefficient is quasilocal in the sense of Axiom 3 in Section \ref{Sect:genDW}. The world function and the van Vleck determinant are similarly quasilocal, so this description holds for $G_{aa'}^\subDW$ as a whole.

Note as well that each of the Hadamard coefficients is symmetric, so
\begin{equation}
        U_{aa'}(x,x') = U_{a'a}(x',x), \quad V_{aa'}(x,x') = V_{a'a} (x',x).
        \label{UVsym}
\end{equation}
This may be argued in various ways. Most simply, the self-adjointness of the differential operator $\delta^b_a \nabla^c \nabla_c - R^{b}{}_{a}$, Stokes' theorem, and the causal properties of the advanced and retarded Green functions imply that $G_{aa'}^\ret = G_{a'a}^\mathrm{adv}$. Eq. \eqref{GretEven} thus implies \eqref{UVsym}, at least for null and timelike-separated points. Symmetry of the Detweiler-Whiting Green function merely requires that this property extend also to spacelike-separated points. Such an extension is argued to be valid in Section 6.4 of \cite{FriedlanderWave}; see also \cite{SymMoretti}.

\subsection{Odd-dimensional propagators}
\label{app:Hadamardodd}

The bitensor $U_{aa'}$ which appears in our odd-dimensional propagator $G_{aa'}^\mathrm{odd}$ is the same as the one which appears in the retarded and advanced Green functions associated with \eqref{Ggreen}. It may be found by factoring out the van Vleck determinant and expanding in the Hadamard series
\begin{equation}
        U_{aa'} = \Delta^{1/2} \sum_{n=0}^\infty \frac{\sigma^n}{n!} \mathsf{U}_{aa'}^{\{ n \}}.
\end{equation}
Unlike its even-dimensional analog \eqref{Uexpand}, the sum here does not necessarily terminate at finite $n$. Nevertheless, the zeroth term in the series is again given by \eqref{U0}, although the odd-dimensional $\alpha_d$ is now computed using \eqref{alphaOdd} instead of \eqref{alphaEven}. Substituting this and \eqref{GretOdd} into \eqref{Ggreen}, the higher-order Hadamard coefficients $\mathsf{U}^{\{n\}}_{aa'}$ may be shown to be the nonsingular solutions to the same transport equations \eqref{Un} which determine the even-dimensional Hadamard coefficients. The $U_{aa'}$ appearing here is again symmetric and quasilocally dependent on the metric, by the same arguments as in the even-dimensional case.

Also note that again, there is no obstacle to working instead with a somewhat-different propagator whose Hadamard series is truncated at finite $n$.

\section{Expansion methods and coincidence limits}
\label{app:CoincidenceLimits}

We now collect various expansion methods and results relevant to the point-particle fields computed in Appendices \ref{app:EvenD} and \ref{app:OddD}.

Many of these expansions involve $\sigma(x,\gamma(\tau))$, Synge's world function specialized to cases in which one argument is evaluated at a specific proper time on a given timelike worldline $\Gamma$. This is assumed to be a smooth function of $x$ and $\tau$, at least if $g_{ab}$ and $\Gamma$ are themselves smooth and $x$ and $\gamma(\tau)$ are sufficiently close. More precisely, we suppose that these points always lie within a convex normal neighborhood. Then, if $\tau$ is varied while $x$ is held fixed near (but not on) $\Gamma$, there exist exactly two ``nearby'' zeros. We call the larger of these the advanced time $\tau_+(x)$ and the smaller the retarded time $\tau_-(x)$. While it is possible to approximate these times in terms of some given coordinate system, we have no need to do so. Instead, we note that Synge's function must factorize via
\begin{equation}
        2 \sigma(x,\gamma(\tau)) = \left[\tau_+(x) - \tau\right]\left[ \tau - \tau_-(x) \right] \Sigma(x,\tau),
        \label{sigExpand}
\end{equation}
where $\Sigma(x,\tau)$ is assumed to be positive and smooth in all regions of interest. For an inertial worldline in flat spacetime, $\Sigma(x,\tau) = 1$. More generally, everything we need is encoded in the various derivatives of $\Sigma(x,\tau)$ evaluated using coincidence limits in which $x \rightarrow \gamma(\tau)$.

It is convenient to also use the retarded and advanced times to introduce the ``radar time''
\begin{align}
        \bar{\tau}(x)\equiv \frac{1}{2}[\tau_{+}(x)+\tau_{-}(x)],
        \label{tBar}
\end{align}
and the ``radar distance''
\begin{equation}
        r(x) \equiv \frac{1}{2} [ \tau_+(x) - \tau_-(x)],
        \label{rDef}
\end{equation}
associated with points $x$ near $\Gamma$. In terms of these functions, \eqref{sigExpand} may be rearranged to read $2\sigma/\Sigma = r^2 - (\tau-\bar{\tau})^2$, from which it follows that
\begin{equation}
        \bar{\tau} = \tau + \frac{\partial}{\partial \tau} \left( \frac{\sigma}{\Sigma} \right), \qquad r^2 = \frac{2 \sigma}{\Sigma} + \left[ \frac{\partial}{\partial \tau} \left( \frac{\sigma}{\Sigma} \right) \right]^2 .
\end{equation}
These expressions imply that if $\sigma/\Sigma$ is smooth, so are $\bar{\tau}$ and $r^2$.

Now solve \eqref{sigExpand} for $\Sigma(x,\tau)$ and consider the substitution $x = \gamma(\tau')$, in which case $\tau_+ = \tau_- = \tau'$:
\begin{align}
        \Sigma(\gamma(\tau'), \tau) = - \frac{ 2 \sigma(\gamma(\tau') , \gamma(\tau) ) }{ (\tau'-\tau)^2 }.
\end{align}
Coincidence limits for the left-hand side or its derivatives follow by evaluating the right-hand side or its derivatives as $\tau' \rightarrow \tau$, using well-known coincidence limits for the derivatives of Synge's function. For example, applying L'H\^{o}pital's rule twice gives
\begin{equation}
        \Sigma(\gamma(\tau), \tau) = - \lim_{\tau' \rightarrow \tau} \dot{\gamma}^{a'}\dot{\gamma}^{b'} \nabla_{a'} \nabla_{b'} \sigma .
\end{equation}
where we have used the vanishing coincidence limits of $\sigma$ and $\nabla_{a'} \sigma$. Further applying
\begin{align}
        \lim_{x' \rightarrow x} \nabla_{a'} \nabla_{b'} \sigma(x,x') = g_{ab},
        \label{coinc1}
\end{align}
it follows that
\begin{align}
        \Sigma(\gamma(\tau),\tau) =1.
\end{align}
Supplementing \eqref{coinc1} by, e.g.,
\begin{equation}
        \lim_{x' \rightarrow x} \nabla_{a'} \nabla_{b} \sigma(x,x') = -g_{ab},
\end{equation}
coincidence limits of $\tau$-derivatives of $\Sigma(x,\tau)$ may be derived similarly. If we specialize to flat spacetime, in which third and higher derivatives of $\sigma$ vanish, it may be shown that
\begin{equation}
\label{SigCoinc}
\begin{gathered}
        \dot{\Sigma} = 0, \quad \ddot{\Sigma} = \frac{1}{6} |\ddot{\gamma}|^2 , \quad \dddot{\Sigma} = \frac{1}{2} (\ddot{\gamma} \cdot \dddot{\gamma}),
        \\
        \Sigma^{(4)}=\frac{1}{15} \left( 8|\dddot{\gamma}|^{2}+9\ddot{\gamma}\cdot\gamma^{(4)}\right),
        \\
        \Sigma^{(5)}=\frac{1}{3}\left( 2\ddot{\gamma}\cdot\gamma^{(5)}+5\dddot{\gamma}\cdot \gamma^{(4)} \right),
        \\
        \Sigma^{(6)}=\frac{1}{28}\left( 20 \ddot{\gamma}\cdot\gamma^{(6)}+64 \dddot{\gamma}\cdot\gamma^{(5)} +45|\gamma^{(4)}|^{2}\right),
\end{gathered}
\end{equation}
when $x = \gamma(\tau)$.

We also need coincidence limits for $\nabla_a \Sigma(x,\tau)$ and its $\tau$ derivatives. First note that differentiating \eqref{sigExpand} with respect to $x$ and rearranging using \eqref{tBar} and \eqref{rDef} implies that
\begin{align}
        \nabla_a \Sigma(x,\tau') = \frac{ 2 \nabla_a\sigma + \nabla_a [ ( \bar{\tau} - \tau')^2 - r^2] \Sigma  }{ (\tau_+ - \tau') ( \tau'- \tau_-) },
        \label{dSigma}
\end{align}
where we have added a prime to the second argument for later convenience. Noting that
\begin{equation}
        \nabla_a \bar{\tau}(\gamma(\tau) ) = - \dot{\gamma}_a (\tau), \qquad \nabla_a r^2(\gamma(\tau) ) = 0,
        \label{gradlimits}
\end{equation}
substituting $x = \gamma(\tau)$ into \eqref{dSigma} gives
\begin{align}
        \nabla_a \Sigma(\gamma(\tau), \tau') =- \frac{2}{(\tau'-\tau)^2 } \Big[ \nabla_a\sigma (\gamma(\tau), \gamma(\tau'))
        \nonumber
        \\
        ~ + (\tau'-\tau) \dot{\gamma}_a(\tau) \Sigma(\gamma(\tau),\tau')   \Big].
\end{align}
Repeatedly applying L'H\^{o}pital's rule to this expression again allows us to evaluate coincidence limits $\tau' \rightarrow \tau$ for $\nabla_a \Sigma(\gamma(\tau),\tau')$ and its $\tau$-derivatives. Specializing to flat spacetime while using \eqref{SigCoinc}, the first few such limits are
\begin{equation}
\begin{gathered}
        \nabla_a \Sigma = \ddot{\gamma}_a, \quad \nabla_a \dot{\Sigma} = \frac{1}{6} (2 \dddot{\gamma}_a - |\ddot{\gamma}|^2 \dot{\gamma}_a ) ,
        \\
        \nabla_a \ddot{\Sigma} = \frac{1}{6} \left[ \gamma^{(4)}_a - 2 (\ddot{\gamma} \cdot \dddot{\gamma} )\dot{\gamma}_a \right],
        \\
        \nabla_{a}\dddot{\Sigma} = \frac{1}{30} \left[  3\gamma^{(5)}_{a}- \left( 8|\dddot{\gamma}|^{2}+9 \ddot{\gamma} \cdot \gamma^{(4)} \right)\dot{\gamma}_{a} \right],
        \\
        \nabla_{a} \Sigma^{(4)} = \frac{1}{15} \left[ \gamma_{a}^{(6)}-2\left( 2 \ddot{\gamma}\cdot\gamma^{(5)} + 5 \dddot{\gamma}\cdot \gamma^{(4)} \right)\dot{\gamma}_{a} \right],
        \\
        \begin{aligned}
        \nabla_{a}\Sigma^{(5)} = \frac{1}{84} \Big[ 4 \gamma_{a}^{(7)} - \Big( 20 \ddot{\gamma}\cdot\gamma^{(6)} + 64 \dddot{\gamma}\cdot\gamma^{(5)}
        \\
        ~ +45|\gamma^{(4)}|^{2} \Big)\dot{\gamma}_{a}\Big].
        \end{aligned}
\end{gathered}
\label{SigCoincGrad}
\end{equation}

Point-particle electromagnetic fields in odd numbers of dimensions are expressed below in terms of coincidence limits of $W_{a}^{ \{ n\} }(x;\lambda)$ and its derivatives, functions defined by (\ref{WDef}) and (\ref{WcoeffDef}). However, we specialize here to flat spacetime, in which case the first of these equations is replaced by \eqref{WDefFlat}. Recalling that $\nabla_b g_{aa'} = \nabla_{b'} g_{aa'} = 0$ in Minkowski spacetimes, the $W_{a}^{ \{ n\} }(x;\lambda)$ can depend only on the particle's worldline and on $\Sigma(x,\tau)$. Using \eqref{SigCoinc} and \eqref{SigCoincGrad}, the first coincidence limits in flat spacetime may be shown to be
\begin{equation}
\label{W}
\begin{gathered}
        W_a^{ \{0\} } = q \dot{\gamma}_a, \qquad W_a^{ \{1\}} = q\ddot{\gamma}_a,
        \\
        W_a^{\{2\}} =q \Big( \dddot{\gamma}_a + \frac{\lambda}{6} |\ddot{\gamma}|^2 \dot{\gamma}_a\Big),
        \\
        W_a^{\{3\}} = q\Big[ \gamma^{(4)}_a + \frac{\lambda}{2} \left( |\ddot{\gamma}|^2 \ddot{\gamma}_a + ( \ddot{\gamma} \cdot \dddot{\gamma} ) \dot{\gamma}_a \right)\Big],
\end{gathered}
\end{equation}
and
\begin{equation}
\label{DW}
\begin{gathered}
        \nabla_{[a} W_{b]}^{\{0\}} = -q ( 1 + \lambda) \dot{\gamma}_{[a} \ddot{\gamma}_{b]},
        \\
        \nabla_{[a} W_{b]}^{\{1\}}  = - q ( 1 + \frac{1}{3} \lambda ) \dot{\gamma}_{[a} \dddot{\gamma}_{b]},
        \\
        \begin{aligned}
        \nabla_{[a} W_{b]}^{\{2\}} = \frac{1}{6} q \Big[ 2 \lambda \ddot{\gamma}_{[a} \dddot{\gamma}_{b]} - \lambda ( 4 + \lambda ) |\ddot{\gamma}|^2 \dot{\gamma}_{[a} \ddot{\gamma}_{b]}
        \\
        - ( 6 + \lambda) \dot{\gamma}_{[a} \gamma^{(4)}_{b]}\Big].
        \end{aligned}
\end{gathered}
\end{equation}

\section{Point-particle fields in even dimensions}
\label{app:EvenD}

This appendix computes various electromagnetic fields associated with monopole point charges in potentially-curved spacetimes for which $d \geq 4$ is even. We start by evaluating the vector potential \eqref{FSDef} for the $S$-field associated with the point-particle current \eqref{Jpoint}. Identifying $G_{aa'}$ with the Detweiler-Whiting Green function \eqref{evenD}, this is more explicitly
\begin{align}
        A_a^\subS (x) = \frac{q}{2} \int \big[ U_{aa'} (x,\gamma(\tau)) \delta^{(d/2-2)}(\sigma(x,\gamma(\tau)))
        \nonumber
        \\
        ~ + V_{aa'}(x,\gamma(\tau)) \Theta(-\sigma(x,\gamma(\tau))) \big] \dot{\gamma}^{a'}(\tau) d\tau.
                \label{eq:ASEven}
\end{align}
The range of $\tau$ values over which this integration is to performed are to be understood as restricted to a normal neighborhood of $x$, in which case the only relevant zeros of $\sigma(x,\gamma(\tau))$ are, for fixed $x$, at $\tau = \tau_\pm (x)$ [cf. \eqref{sigExpand}]. Hence,
\begin{align}
        \label{eq:SingFieldEven}
        A_a^\subS = \frac{q}{2} \Bigg[ \sum_{\tau \in \{ \tau_\pm \} } \frac{1}{|\dot{\sigma}|} \left( -\frac{\partial}{\partial \tau}\frac{1}{\dot{\sigma}} \right)^{d/2-2} U_{aa'}\dot{\gamma}^{a'}
        \nonumber
        \\
         ~+ \int_{\tau_-}^{\tau_+} V_{aa'} \dot{\gamma}^{a'}  d\tau \Bigg],
\end{align}
where $\dot{\sigma} = \partial \sigma(x,\gamma(\tau))/\partial \tau$ and the sum denotes that one is to substitute $\tau = \tau_+$ and then add to that the same expression evaluated at $\tau = \tau_-$. While other null geodesics may exist between the particle's worldline and $x$, it is only the ``closest two'' which are included in this expression.

The retarded Green function may be shown to have to have the form \eqref{GretEven} at least within a normal neighborhood, so the retarded vector potential is
\begin{align}
        \label{eq:RetFieldEven}
        A_a^\ret = q \Bigg[ \left. \frac{1}{|\dot{\sigma}|} \left( -\frac{\partial}{\partial \tau}\frac{1}{\dot{\sigma}} \right)^{d/2-2} U_{aa'}\dot{\gamma}^{a'} \right|_{\tau = \tau_-}
        \nonumber
        \\
         ~+ \lim_{\epsilon \to 0^+} \int^{\tau_--\epsilon}_{-\infty} G^\ret_{aa'} \dot{\gamma}^{a'}  d\tau \Bigg].
\end{align}
The Green function in the second line here is left as-is to allow for integrations beyond the normal neighborhood, in which case the Hadamard form \eqref{GretEven} can fail to remain valid.

If the full electromagnetic field $F_{ab}$ is identified with the retarded field $F^\ret_{ab} = 2 \nabla_{[a} A_{b]}^\ret$, it follows from \eqref{Fhat} that $\hat{F}_{ab} = 2 \nabla_{[a} ( A^\ret_{b]} - A^\subS_{b]})$. In Minkowski spacetime, this is equivalent to what is often called the radiative field, one-half of the retarded minus advanced fields. In more general spacetimes, a vector potential for $\hat{F}_{ab}$ with retarded boundary conditions may be written as the difference between (\ref{eq:RetFieldEven}) and (\ref{eq:SingFieldEven}):
\begin{align}
        \label{eq:HatFieldEven}
        \hat{A}_a &= q \Bigg[ \left. \frac{1}{2|\dot{\sigma}|} \left( -\frac{\partial}{\partial \tau}\frac{1}{\dot{\sigma}} \right)^{d/2-2} U_{aa'}\dot{\gamma}^{a'} \right|^{\tau = \tau_-}_{\tau = \tau_+}
        \nonumber
        \\
        & ~- \frac{1}{2} \int_{\tau_-}^{\tau_+} V_{aa'}\dot{\gamma}^{a'} d \tau + \lim_{\epsilon \to 0^+} \int^{\tau_- - \epsilon}_{-\infty} G^\ret_{aa'} \dot{\gamma}^{a'}  d\tau \Bigg].
\end{align}
Although it is not obvious from this expression, the effective field is finite, and indeed smooth, even on the worldline, essentially because it satisfies the source-free Maxwell equations.

\section{Point-particle fields in odd dimensions}
\label{app:ppOdd}

We now compute point-particle fields in odd numbers of dimensions. Section \ref{app:Sfield} starts by obtaining a vector potential $A_a^\subS$ for the $S$-field $F_{ab}^\subS$ associated with the point-particle current \eqref{Jpoint}, identifying the propagator $G_{aa'}$ by which these fields are defined with the $G^\mathrm{odd}_{aa'}$ given in \eqref{EqLogProp}. The final result, summarized by \eqref{ASfin}, is a series involving the radar distance $r$ away from the particle's worldline $\Gamma$ [as defined by \eqref{rDef}]. This series involves positive and negative powers of $r$, $\ln r$, and coefficients which can depend smoothly on $x$.

Next, the point-particle retarded field $A_a^\ret(x)$ is computed in Section \ref{app:Retfield}, again as a series involving $r$. Both the retarded and $S$ fields diverge on $\Gamma$, although we show in Section \ref{app:Hatfield} that their difference is smooth. This last result essentially constitutes our verification that Axiom 4 of Section \ref{Sect:genDW} is satisfied by $G_{aa'}^\mathrm{odd}$.

\label{app:OddD}

\subsection{The $S$-field}

\label{app:Sfield}

The point-particle $S$-field vector potential may be found by evaluating \eqref{FSDef} with $G_{aa'} = G_{aa'}^\mathrm{odd}$  and $J^a = J^a_\mathrm{pp}$. Given the form \eqref{EqLogProp0} for $G_{aa'}^\mathrm{odd}$, it is useful to introduce the auxiliary family of propagators
\begin{equation}
        \tilde{G}_{aa'}(x,x';\lambda) \equiv U_{aa'}(x,x') [2 \sigma(x,x')]^\lambda\Theta(\sigma(x,x')),
\end{equation}
and the associated point-particle fields
\begin{equation}
        \tilde{A}_a(x;\lambda) = q \int_{\tau_-(x)}^{\tau_+(x)} \!\! [2 \sigma(x,\gamma(\tau))]^\lambda U_{aa'}(x,\gamma(\tau)) \dot{\gamma}^{a'}(\tau) d\tau.
        \label{Aplus}
\end{equation}
Once this potential is known, the point-particle $S$-field follows from
\begin{align}
        A_a^\subS = \frac{(-1)^{\frac{1}{2}-\lambda_d}}{2\pi}  \lim_{\lambda \rightarrow \lambda_d} \ell^{2\lambda} \frac{\partial}{\partial \lambda} ( \ell^{-2\lambda} \tilde{A}_a ),
\label{ASDef}
\end{align}
where $\ell > 0$ is the arbitrary lengthscale used in the construction of $G_{aa'}^\mathrm{odd}$, the dimension-dependent number $\lambda_d$ is given by \eqref{lambdaDef}, and the limit implies an analytic continuation in $\lambda$.

A series expansion for $\tilde{A}_a$ may now be found by substituting the factorization \eqref{sigExpand} for $\sigma$ into \eqref{Aplus}. Doing so results in
\begin{align}
        \label{Aplus2}
        \tilde{A}_a (x;\lambda) = \alpha_d \int_{\tau_-(x)}^{\tau_+(x)}\!\! d\tau [\tau_+(x) - \tau]^\lambda [\tau- \tau_-(x)]^\lambda
        \nonumber
        \\
        ~ \times W_a(x,\tau;\lambda),
\end{align}
where $\alpha_d$ is given by \eqref{alphaOdd} and it convenient to define
\begin{align}
        W_a(x,\tau;\lambda) \equiv \frac{q}{\alpha_d} U_{aa'}(x,\gamma(\tau)) \dot{\gamma}^{a'} (\tau) \Sigma^\lambda(x, \tau).
        \label{WDef}
\end{align}
Expanding $W_a(x,\tau;\lambda)$ about $\tau = \bar{\tau}(x)$ as defined in \eqref{tBar}, we find that
\begin{align}
        W_a(x,\tau;\lambda) &= \sum_{n=0}^\infty \frac{1}{n!} [\tau- \bar{\tau}(x)]^n W_a^{ \{ n \} }(x;\lambda),
\end{align}
in terms of the coefficients
\begin{align}
         W_a^{ \{ n \} } (x;\lambda)\equiv \left. \frac{\partial^{n}}{\partial \tau^{n}} W_a (x,\tau;\lambda) \right|_{\tau = \bar{\tau}(x)}.
        \label{WcoeffDef}
\end{align}
Substituting these expressions into \eqref{Aplus2} now yields
\begin{align}
        \tilde{A}_a
        = 2 \alpha_{d} \sum_{n=0}^\infty \frac{ W^{ \{ 2n \} }_a }{ (2n)! } \int_{\bar{\tau}}^{\tau_+} \!\! d\tau (\tau-\bar{\tau})^{2(n+\lambda)} \nonumber\\
        \times\left[ \left( \frac{ r }{\tau-\bar{\tau}} \right)^2 -1 \right]^\lambda.
        \label{AplusInter}
\end{align}
If $\lambda>-1$, the integral on the right-hand side is well-defined and
\begin{equation}
        \tilde{A}_a =  \alpha_d \sum_{n=0}^\infty \frac{\Gamma(n+\frac{1}{2}) \Gamma(\lambda+1)}{ (2n)! \Gamma(n+\lambda+\frac{3}{2})} W_a^{\{2n\}} r^{1+2 (n+\lambda)}.
        \label{Atilde}
\end{equation}
However, it follows from \eqref{lambdaDef} that $\lambda_d < -1$ in five or more dimensions. The analytic continuation associated with the limit in \eqref{ASDef} nevertheless implies that the right-hand side of $\eqref{Atilde}$ remains valid as long as it may be analytically continued to $\lambda \to \lambda_d$.

Carrying out this continuation, $\Gamma(n + \lambda + \frac{3}{2})$ diverges for all $n \leq \frac{1}{2} (d-5)$. Such terms therefore go to zero in the sum \eqref{Atilde} and
\begin{align}
        \lim_{\lambda \rightarrow \lambda_d} \tilde{A}_a (x;\lambda) =  \alpha_{d} \!\! \sum_{n=\frac{1}{2}(d-3)}^\infty \frac{\Gamma(n + \frac{1}{2} ) \Gamma(2-\frac{1}{2} d)}{(2n)! \Gamma(n+\frac{1}{2}(5-d))}\nonumber\\
        ~ \times W_a^{\{ 2n \} }(x;\lambda_d) r^{2n - (d-3)}.
        \label{AtildeLim}
\end{align}
This depends only on non-negative even powers of $r(x)$ and on the smooth functions $W^{\{2n\}}_a (x;\lambda_d)$; the overall result is smooth near the particle's worldline.

Computing $A^\subS_a$ requires not only $\tilde{A}_a$ continued to the appropriate value of $\lambda$, but also a continuation for the $\lambda$-derivative of that field. Differentiating \eqref{Atilde}, one finds that that is
\begin{align}
        \partial_\lambda \tilde{A}_a =\alpha_{d} \sum_{n=0}^\infty \frac{ \Gamma(n+\frac{1}{2}) \Gamma(\lambda+1) }{ (2n)! \Gamma(n + \lambda + \frac{3}{2} ) } r^{1+2(n+\lambda)}  \Big[ \partial_\lambda W_a^{\{2n\}}
        \nonumber
        \\
        ~ + \Big( H_\lambda - H_{n+\lambda+\frac{1}{2}} + 2 \ln r \Big) W_a^{\{2n\}} \Big]
\end{align}
for general values of $\lambda$, where $H_\mu$ denotes the $\mu$th harmonic number. Taking the $\lambda \rightarrow \lambda_d$ limit here requires some care since the factor of $\Gamma(n+\lambda+\frac{3}{2})$ in the denominator and the harmonic number $H_{n+\lambda+\frac{1}{2}}$ in the numerator both diverge in that limit, for all $n \leq \frac{1}{2} (d-5)$. What is important however is the limit of their ratio, which may be shown to be
\begin{align}
        \lim_{\lambda \rightarrow \lambda_d} \frac{ H_{n+\lambda+\frac{1}{2}} }{ \Gamma(n+\lambda+\frac{3}{2}) } &= (-1)^{\frac{1}{2}(d-3)-n} \Gamma( \tfrac{1}{2} (d-3) - n)
\end{align}
for all $n \leq \frac{1}{2} (d-5)$. Hence,
\begin{widetext}
\begin{align}
        \lim_{\lambda \rightarrow \lambda_d} \partial_\lambda \tilde{A}_a =  \alpha_d \Bigg\{ \sum_{ n=\frac{1}{2}(d-3) }^\infty \! \! \frac{ \Gamma(n+\frac{1}{2}) \Gamma(2-\frac{d}{2}) }{ (2n)! \Gamma(n + \frac{1}{2} (5-d) ) } \Big[\left( H_{1-\frac{d}{2}} - H_{n-\frac{1}{2}(d-3)} +  2 \ln r \right) W_a^{\{2n\}} + \partial_\lambda W_a^{\{2n\}}\Big] r^{2n-(d-3)}
        \nonumber
        \\
        ~  - (-1)^{\frac{1}{2}(d-3)} \sum_{n=0}^{\frac{1}{2}(d-5)} \frac{ (-1)^n \Gamma(n+\frac{1}{2}) \Gamma(2-\frac{d}{2}) \Gamma(\frac{1}{2}(d-3)-n) }{ (2n)! } W_a^{\{2n\}} r^{2n-(d-3)} \Bigg\}.
\end{align}
\end{widetext}
The full $S$-field is found by substituting this equation and \eqref{AtildeLim} into \eqref{ASDef}. The result is \eqref{ASfin} in the main text.

\subsection{The retarded field}
\label{app:Retfield}

We now derive the retarded point-particle field, an expansion of which may be found using an integral analogous to \eqref{AplusInter}. Unfortunately, the relevant integration is no longer performed over the interval $\tau \in (\tau_-, \tau_+)$, but instead runs over all $\tau < \tau_-$. There are various reasons for which it is undesirable to attempt expansions over this infinite domain, so we initially consider integrals for the retarded vector potential which are truncated at some finite time $T < \tau_-$. We eventually find it convenient to let $T$ be only slightly less than $\tau_-$, although it may be viewed more generally for now.

Convolving the odd-dimensional retarded Green function \eqref{GretOdd} with the point-particle current density \eqref{Jpoint} while using the expansion coefficients defined by \eqref{WcoeffDef}, the appropriate truncated field can be shown to be
\begin{align}
        A_a^T = \alpha_{d} \sum_{n=0}^\infty \frac{ (-1)^n }{n!} W_a^{ \{ n \} } \int_T^{\tau_-} d\tau (\bar{\tau}-\tau)^{n+2\lambda}
        \nonumber
        \\
        ~ \times \left[ 1 - \left( \frac{ r }{ \bar{\tau} - \tau} \right)^2 \right]^{\lambda},
        \label{ATinter1}
\end{align}
where we have omitted the implicit limit $\lambda \rightarrow \lambda_d$. From this, the full retarded field strength follows via
\begin{equation}
        F_{ab}^\ret = 2 \nabla_{[a} A_{b]}^T + 2 q \int^T_{-\infty} \! \!  \nabla_{[a} G_{b]b'}^\ret \dot{\gamma}^{b'} d\tau.
        \label{FretFromTrunc}
\end{equation}
We choose to consider $F_{ab}^\ret$ here instead of $A_a^\ret$ in order to avoid convergence problems when $d=3$.

Now, the truncated vector potential may be evaluated by applying the binomial theorem to expand the term in square brackets in \eqref{ATinter1}, giving
\begin{align}
        A_a^T = \alpha_d \sum_{n=0}^\infty \sum_{k=0}^\infty \frac{ (-1)^{n+k} \Gamma(1+\lambda) }{n! k! \Gamma(1+\lambda-k) } W_a^{ \{ n \} } r^{2k} \nonumber\\
        \times\int_T^{\bar{\tau} - r} \! \!  d\tau (\bar{\tau}-\tau)^{n+2(\lambda-k)} .
\end{align}
Evaluating the $\lambda \rightarrow \lambda_d$ limit of this expression requires some care. Omitting details, the result is that
\begin{widetext}
\begin{align}
        \label{eq:AT}
        A_a^T = \alpha_{d} \Bigg\{ \sum_{n = 0}^\infty \sum_{k \neq k_n }^\infty \frac{ (-1)^{k+n} \Gamma(2-\frac{d}{2}) (\bar{\tau}-T)^{n+3-d-2k}  W_a^{ \{ n \} } r^{2k} }{n! k! (n+3-d-2k) \Gamma(2-\frac{d}{2} -k)  } + \sum_{n=0}^{\frac{1}{2}(d-5)} \frac{ \Gamma(2-\frac{d}{2} ) \Gamma(\frac{1}{2}(d-3)-n) W_a^{ \{ 2n \} } }{ 2  (2n)! \Gamma(\frac{1}{2}-n)  r^{(d-3) -2n} }
        \nonumber
        \\
        ~ -  \sum_{n=\frac{1}{2} (d-3)}^\infty \frac{ (-1)^{ n+ \frac{1}{2}(d-3)} \Gamma(2-\frac{d}{2}) r^{2n + 3 - d} }{2 (2n)! \Gamma(\frac{1}{2}-n) \Gamma(n-\frac{1}{2}(d-5)) } \left[  H_{-\frac{1}{2}-n} - H_{n-\frac{1}{2}(d-3)} + 2 \ln \left( \frac{ r }{ \bar{\tau} - T} \right) \right] W_a^{ \{ 2p \} } \Bigg\},
\end{align}
where $k_n \equiv \frac{1}{2}[n- (d-3)]$. This can be substituted into \eqref{FretFromTrunc} to obtain the full (non-truncated) retarded field for a point particle in an odd-dimensional spacetime.

\subsection{The effective field}
\label{app:Hatfield}

Our final task in this appendix is to compute the point-particle effective field with retarded boundary conditions in odd dimensions. Defining the effective cut-off potential by $\hat{A}_a^T \equiv A_a^T - A_a^S$ and comparing \eqref{eq:AT} with \eqref{ASfin}, all logarithms and negative powers of $r$ exactly cancel, leaving
\begin{align}
        \hat{A}_a^T = \alpha_d \Bigg\{ \sum_{n = 0}^\infty \sum_{k \neq k_n }^\infty \frac{ (-1)^{k+n} \Gamma(2-\tfrac{d}{2}) (\bar{\tau}-T)^{n+3-d-2k} W_a^{ \{ n \} }  r^{2k} }{n! k! (n+3-d-2k) \Gamma(2-\frac{d}{2} -k) }  + \! \! \sum_{n= \frac{1}{2} (d-3) }^\infty \! \frac{ (-1)^{ n+ \frac{1}{2}(d-5)} \Gamma(2-\tfrac{d}{2})  r^{2n-(d-3)}  }{2 (2n)! \Gamma(\frac{1}{2}-n) \Gamma(n-\frac{1}{2}(d-5)) }
        \nonumber
        \\
        ~ \times \left[ \left( H_{-\frac{1}{2}-n} - H_{1-\frac{d}{2}} - 2 \ln \left( (\bar{\tau} - T )/\ell  \right) \right) W_a^{ \{ 2n \} } - \partial_\lambda W_a^{ \{ 2n \} } \right]    \Bigg\}.
        \label{AhatT}
\end{align}
This depends on $x$ only via non-negative even powers of $r(x)$, the smooth coefficients $W_a^{ \{n\} }(x;\lambda_d)$, and $\bar{\tau}(x)$. Moreover, it follows from \eqref{Fhat} and \eqref{FretFromTrunc} that the full effective field strength with retarded boundary conditions is
\begin{equation}
        \hat{F}_{ab} = 2 \nabla_{[a} \hat{A}_{b]}^T +  2 q \int^T_{-\infty} \!\! \nabla_{[a} G_{b]b'}^\ret \dot{\gamma}^{b'} d\tau.
        \label{FhatT}
\end{equation}
As long as the series in \eqref{AhatT} converge, any number of derivatives of the effective field exist, even on $\Gamma$, because $\bar{\tau}(x) - T \neq 0$ everywhere of interest and $\bar{\tau}(x)$ and $r^2(x)$ are smooth; see Appendix \ref{app:CoincidenceLimits}.

Evaluating the leading-order self-force and self-torque acting on a point particle requires that we evaluate $\hat{F}_{ab}(x)$ on the particle's worldline, where $r \to 0$. Discarding terms in the truncated potential \eqref{AhatT} which are $\mathcal{O}(r^{2})$, we find that
\begin{align}
        \hat{A}_{a}^T = \alpha_{d} \left\{ \sum_{n\neq d-3}^\infty \frac{ (T-\bar{\tau})^{n-(d-3)}}{ n! (n+3-d) } W^{ \{ n \} }_a + \frac{1}{(d-3)!} \left[ W_a^{ \{ d-3 \} } \ln\left( (\bar{\tau} - T )/ \ell \right) + \frac{1}{2} \partial_\lambda W_a^{ \{ d-3 \} } \right] \right\} .
\end{align}
Using this in \eqref{FhatT} and letting $r \to 0^+$, individual terms in the resulting expression for $\hat{F}_{ab}(\gamma(\tau))$ depend on the arbitrarily-chosen cutoff time $T$. Nevertheless, all such terms taken together cannot depend on $T$. We are therefore free to choose $T = \tau - \epsilon$ for some $\epsilon > 0$, and then to take the limit $\epsilon \rightarrow 0^+$. Doing so eliminates the infinite sum in $n$, leaving only
\begin{align}
        \hat{F}_{ab}(\gamma(\tau))  = 2 \lim_{\epsilon \to 0^+} \Bigg\{ q \int^{\tau-\epsilon}_{-\infty} \nabla_{[a} G^\ret_{b]b'} \dot{\gamma}^{b'} d\tau' - \alpha_d \Bigg[ \sum_{n=0}^{d-4} \frac{ (-1)^n }{ n! } \left( \frac{ \nabla_{[a} W_{b]}^{ \{ n \} } }{ d - 3- n} + \frac{1}{\epsilon } \dot{\gamma}_{[a} W_{b]}^{ \{n \} } \right) \frac{1}{\epsilon^{d-3-n}}
        \nonumber
        \\
        ~ + \frac{1}{(d-3)!} \left( \frac{1}{\epsilon} \dot{\gamma}_{[a} W_{b]}^{ \{ d-3 \} } - \nabla_{[a} W_{b]}^{ \{ d-3 \} } \ln (\epsilon/\ell) - \frac{1}{2} \partial_\lambda \nabla_{[a} W_{b]}^{ \{ d-3 \} } - \frac{1}{(d-2)} \dot{\gamma}_{[a} W_{b]}^{ \{ d-2 \} } \right) \Bigg] \Bigg\} .
        \label{FhatGenFin}
\end{align}
\end{widetext}
A version of this expression specialized to flat spacetime is given by \eqref{FhatFin} in the main text. In either form, the coefficients $W_a^{ \{ n \} }(\gamma(\tau);\lambda)$ and $\nabla_{[a} W_{b]}^{ \{ n \} }(\gamma(\tau);\lambda)$ which appear here are to be evaluated in their coincidence limits, and it is implicit that $\lambda = \lambda_d = 1-d/2$. The first four undifferentiated and the first three differentiated coefficients of this kind are given explicitly in flat spacetime by \eqref{W} and \eqref{DW}. These are sufficient to determine $\hat{F}_{ab}$ in full for $d=3$ and $d=5$. Higher-dimensional results follow by extending the limit calculations described in Appendix \ref{app:CoincidenceLimits}.

\bibliographystyle{apsrev4-1}
\bibliography{selfforce}

\end{document}